\documentclass[a4paper,11pt]{article}
\pdfoutput=1 
\usepackage{jcappub} 
\usepackage[T1]{fontenc}
\usepackage{mwe}
\usepackage{subfig}
\usepackage{xcolor}
\usepackage{floatrow}
\usepackage{multirow}
\usepackage{comment}

\newcommand{\SAcomm}[1]{\textbf{{\color{blue} [Shaz: #1]}}}

\newcommand{\eV}{\mathrm{eV}}

\newcommand{\taur}{\tau_\mathrm{reio}}
\newcommand{\taudm}{\tau_\mathrm{DCDM}}
\newcommand{\Gammadm}{\Gamma_\mathrm{DCDM}}
\newcommand{\fdm}{f_\mathrm{DCDM}}
\newcommand{\fL}{f_L}

\providecommand{\sorthelp}[1]{}

\title{\boldmath Do you smell something decaying? Updated linear constraints on decaying dark matter scenarios}

\author[a,b]{S. Alvi,}
\author[a,b]{T. Brinckmann,}
\author[b]{M. Gerbino,}
\author[b]{M. Lattanzi,}
\author[a,b,c]{and L. Pagano}

\affiliation[a]{Dipartimento di Fisica e Scienze della Terra, Universit\`a degli Studi di Ferrara, via Giuseppe Saragat 1, I-44122 Ferrara, Italy}
\affiliation[b]{Istituto Nazionale di Fisica Nucleare, Sezione di Ferrara, via Giuseppe Saragat 1, I-44122 Ferrara, Italy}
\affiliation[c]{Institut d'Astrophysique Spatiale, CNRS, Univ. Paris-Sud, Universit\'{e} Paris-Saclay, B\^{a}t. 121, 91405 Orsay cedex, France}

\emailAdd{shahbaz.nihalalvi@gmail.com}
\emailAdd{thejs.brinckmann@gmail.com}
\emailAdd{gerbinom@fe.infn.it}
\emailAdd{lattanzi@fe.infn.it}
\emailAdd{luca.pagano@unife.it}

\abstract{The stability of particles in the cosmic soup is an important property that can affect the cosmic evolution. In this work, we update the constraints on the decaying cold dark matter scenario, when the decay products are effectively massless. We assume, as a base case, that all of dark matter is unstable and it can decay on cosmological time scales. We then extend the analysis to include the scenario where only a fraction of dark matter is unstable, while the remaining part is composed of the standard, stable, dark matter. We consider observations of cosmological probes at linear scales, i.e., Planck 2018 cosmic microwave background temperature, polarization, and lensing measurements, along with geometrical information from baryon acoustic oscillation (BAO) measurements from SDSS DR7, BOSS DR12, eBOSS DR16 and 6dFGS, to derive conservative constraints on the dark matter decay rate. We consider these dataset separately, to asses the relative constraining power of each dataset, as well as together to asses the joint constraints. We find the most stringent upper limit on the decay rate of decaying cold dark matter particles to be $\Gammadm < 0.129 \times 10^{-18}\,\mathrm{s}^{-1}$ (or, equivalently, the dark matter lifetime $\taudm > 246$ Gyr) at 95\% C.L. for the combination of Planck primary anisotropies, lensing and BAO. We further explore one-parameter extensions of our baseline DCDM model. Namely, we vary the sum of neutrino masses, the curvature density parameter, and the tensor-to-scalar ratio along with the DCDM parameters. When varying the tensor-to-scalar ratio we also add data from the BICEP/Keck experiment.}

\begin{document}
\maketitle
\flushbottom

\section{Introduction}
\label{Intro}
Dark matter (DM) is an essential component of our understanding of the dynamics of the early universe and galaxy formation. Despite the overwhelming evidence for its existence, the nature of the DM is still shrouded in mystery. Efforts to identify the existence of DM have primarily relied on its gravitational interaction with light and ordinary matter.

Considering the role of DM in structure formation, the DM particle must possess a high degree of stability, perhaps even absolute, as it must fulfill its role in providing the gravitational potential for baryonic matter to form galaxies and galaxy clusters. In the standard model of particle physics, however, most of the particles are unstable having decay lifetime spanning several orders of magnitude; absolute stability of a particle species is enforced through some form of exact symmetry. Thus it would be presumptuous to say that DM particle possesses absolute stability.

In this paper, we will explore the possibility that the DM is long-lived but unstable on cosmological timescales. In particular, we assume that DM is cold and that it decays to some form of dark radiation.

Considering only ``invisible'' decays, as opposed to ``visible'' decays to electromagnetically-interacting particles like photons or electrons, is a conservative choice when seeking to put constraints on the DM lifetime. Indeed, visible decays would have a larger observable effect for a given decay rate and are strongly constrained, for example by measurements of the ionization state of the intergalactic medium \cite{Yeung:2012ya,Liu:2016cnk,Slatyer:2016qyl,Oldengott:2016yjc,Poulin:2016anj,Stocker:2018avm}.
We consider both possibilities of all the dark matter, or only a fraction of it, being unstable.

Several particle physics scenarios might justify a DM component that mostly, or exclusively, decays into dark radiation (DR).
A possibility is simply that both the DM and the DR belong to a dark sector that is completely decoupled from the standard model, and thus interacts only gravitationally with it, such as in the string landscape \cite{Halverson:2018vbo}. Alternatively, it might be that the dark sector couples to the standard model only through neutrinos. This possibility is realized in scenarios of neutrino mass generation in which neutrino masses arise from the coupling of the Majoron, the Nambu-Goldstone boson of broken lepton number, to Majorana neutrinos \cite{Chikashige:1980ui,Gelmini:1980re,Schechter:1981cv}. In this scenario, either the Majoron itself or one of the heavy sterile neutrinos might play the role of DM.

Starting from the neutrino decay in~\cite{Adams:1998nr}, various attempts have been made to study the effects of invisible  DM decays on the cosmological evolution and to constrain the decay rate $\Gammadm$ (or equivalently the lifetime $\taudm$) of a decaying cold dark matter (DCDM) particle. An early investigation of DM decaying into relativistic products was reported in \cite{Kaplinghat:1999xy}, improving upon~\cite{Lopez:1998jt}, where the lower limit on decay lifetime was found to be $10^{12}$ seconds ($\taudm > 32$~kyrs) using measurements of the cosmic microwave background (CMB) available at the time. The limits dramatically improved using WMAP measurements of the CMB anisotropies \cite{Ichiki:2004vi,Lattanzi:2007ux,Lattanzi:2013uza}; for example, Ref.~\cite{Lattanzi:2013uza} found $\taudm > 50$~Gyr using the final, 9-year WMAP data release. In~\cite{DeLopeAmigo:2009dc}, the authors additionally included in their analysis Type-Ia supernovae, weak lensing, Lyman-$\alpha$ measurements, along with CMB, finding a lower limit of $\taudm \gtrsim$ 100 Gyr  on the lifetime of dark matter.

More recently, significantly stronger constraints were reported in \cite{Audren:2014bca} and 
\cite{Poulin:2016nat} using Planck 2013 and 2015 CMB data, respectively, as well as large-scale structure (LSS) data (baryon acoustic oscillations -- BAO -- from BOSS DR11 and galaxy power spectrum from WiggleZ).
Ref. \cite{Audren:2014bca} also included 9-year WMAP and BICEP2 measurements and derived a lower limit on the DM lifetime of $\taudm > 160$ Gyr and 200 Gyr with and without BICEP2 data, respectively, in a cosmological model with non-vanishing primordial tensor perturbations. 

While previous analyses assumed that all of the DM is decaying (Ref.~\cite{Ichiki:2004vi} being a notable exception), Ref.~\cite{Poulin:2016nat} explored the possibility that only a fraction, $\fdm$, of the DM is unstable. The authors found from CMB data that the product $\fdm\, \Gammadm$ should be $< 6.3 \times 10^{-3}$ Gyr$^{-1}$, or $\taudm/\fdm > 159\,\mathrm{Gyr}$ ($\fdm\, \Gammadm < 5.9 \times 10^{-3}$ Gyr$^{-1}$ or $\taudm/\fdm > 170\,\mathrm{Gyr}$ when including LSS). Ref.~\cite{Xiao:2019ccl} subsequently considered the impact of redshift-space distortion and kinetic Sunyaev-Zeldovich measurements on constraints on fractional decay scenarios. Ref.~\cite{Nygaard:2020sow} also considered cases where only a fraction of the DM is unstable and updated constraints for Planck 2018 CMB data with and without BAO data from BOSS DR12, finding for their very long lived regime $\fdm\, \Gammadm < 4.01 \times 10^{-3}$ Gyr$^{-1}$ ($\taudm/\fdm > 249\,\mathrm{Gyr}$) and $\fdm\, \Gammadm < 3.72 \times 10^{-3}$ Gyr$^{-1}$ ($\taudm/\fdm > 269\,\mathrm{Gyr}$), respectively.

The effects of DCDM on the nonlinear growth of structure have been explored in Refs. \cite{Enqvist:2015ara,Enqvist:2019tsa,Dakin:2019dxu,Hubert:2021khy} through N-body simulations. Ref.~\cite{Enqvist:2015ara} reports lower bounds on the DM lifetime to be $\taudm > 97$ Gyr by combining Planck 2013 and 9-year WMAP CMB data with weak lensing data from CFHTLenS. Notably, this is weaker than their CMB data plus SDSS DR9 and 6dFGS BAO bound of $\taudm > 140$ Gyr. An updated analysis by the same authors, using Planck2015 results (including cluster counts from observations of the Sunyaev-Zel'dovich effect) combined with the KiDS450 and the measurements of the baryon acoustic scale yields $\taudm > 175$ Gyr \cite{Enqvist:2019tsa}. Recently, the effect of DM decays on mildly nonlinear scales has been studied in Ref.~\cite{Simon:2022ftd} using the Effective Field Theory of Large Scale Structures formalism, reporting a lower bound on the DM lifetime $\taudm/\fdm > 250$ Gyr.

Other models of decaying DM have been explored in the literature. For example, the possibility that the decay products are not massless has been considered e.g. in Refs.~\cite{Aoyama:2011ba,Aoyama:2014tga,Blackadder:2014wpa,Blackadder:2015uta,Haridasu:2020xaa,Abellan:2020pmw,Abellan:2021bpx,Davari:2022uwd}. Ref. \cite{Blinov:2020uvz} instead studies the case in which the unstable DM is warm. Both the basic DCDM model and the variations just mentioned have recently sparked interest as possible solutions of cosmological tensions (see e.g. Refs.~\cite{Blackadder:2014wpa,Berezhiani:2015yta,Enqvist:2015ara,Bringmann:2018jpr,Pandey:2019plg,Haridasu:2020xaa,Blinov:2020uvz,Abellan:2020pmw,Nygaard:2020sow,Abellan:2021bpx,Davari:2022uwd} or see Refs.~\cite{Verde:2019ivm,DiValentino:2021izs,Schoneberg:2021qvd} for reviews).

In this paper, we derive constraints on the DM lifetime, exploring both the possibility that all of the DM or only a fraction of it is unstable, from the impact of the decay at the background and linear perturbations level. The constraints presented in this manuscript can be regarded as conservative, thanks to the combination of the data we used and the invisible decay channel we considered in this work. 

The paper is structured as follows. 
In Section \ref{sec:eff_cosmo_obv}, we outline the effects of DCDM on the considered cosmological observables, namely the CMB power spectra and the BAO measurements. In Section \ref{sec:data_param_est}, we report the results of our analysis considering both the scenario in which all of dark matter is composed of a single unstable species, and the one in which only a fraction of the total dark matter is unstable. In Section \ref{sec:data_param_est} we also report our findings from extended DCDM models. Finally, in Sections \ref{sec:particle_physics} and \ref{sec:conclusions}, we discuss the implications of our findings on particle physics models and we draw the final conclusions. 

\section{Effects on cosmological observables}
\label{sec:eff_cosmo_obv}

A given DCDM model is specified by fixing the dark matter abundance and decay rate. A particularly convenient way to parameterize the dark matter abundance is to use its value at early times (i.e., at $t\ll \taudm$), long before decays kick off, as done in \cite{Lattanzi:2007ux}. With such a choice, changing the decay rate $\Gammadm$ only affects the late-time cosmology.  
In this paper, as a proxy of the DCDM energy density at early times, we use $\omega^\text{DCDM}_\text{ini}$, defined as \cite{Audren:2014bca}:
\begin{equation}
\omega^\text{DCDM}_\text{ini} \equiv \Omega_\mathrm{DCDM} h^2 \exp(\Gammadm t_0) = \frac{\left(\rho_\mathrm{DCDM}a^3\right)_{t\ll \taudm}}{\rho_{c,0}} h^2 \, .
\label{eq:omega_dcdm}
\end{equation}
In the case of vanishing value of decay rate (stable dark matter), $\omega^\text{DCDM}_\text{ini}$ would be equal to the present day density of standard, stable, cold dark matter.
For what concerns the decay rate $\Gammadm$, in the following we will often present our results in terms of $\Gamma_{18} = \Gammadm/10^{-18}\,\mathrm{s}^{-1}$, namely
the decay rate in units of $10^{-18}\,\mathrm{s}^{-1}$. In this parametrization, $\Gamma_{18} = 1.0$ roughly corresponds to a dark matter lifetime $\taudm \approx 32$~Gyr.
In order to compute cosmological observables in the presence of DCDM, we use \texttt{CLASS v2.9}~\cite{Blas:2011rf,Lesgourgues:2011re,Lesgourgues:2011rh} as our Boltzmann solver.

The theoretical framework of the DCDM and its effects on cosmological quantities have been discussed in detail in  the literature \cite{Kaplinghat:1999xy,Lattanzi:2007ux,Audren:2014bca,Poulin:2016nat}. Here we provide a short summary that we complement with the aid of relevant figures. In most of the figures, we report the ratio between relevant cosmological quantities computed for different values of the DM decay rate and the corresponding quantities computed for $\Gamma_{18}=0$. The remaining cosmological parameters are kept fixed to the same fiducial values. We note that, for illustrative purposes, the choice of the values of $\Gamma_{18}$ are exaggerated if compared to those currently allowed. In some figures, we also report current measurements to put the comparison in context with current sensitivity from cosmological surveys.

\begin{figure}
\begin{minipage}{.5\linewidth}
\centering
\subfloat[]{\label{fig:cosmo_quant:Hz}\includegraphics[scale=.35]{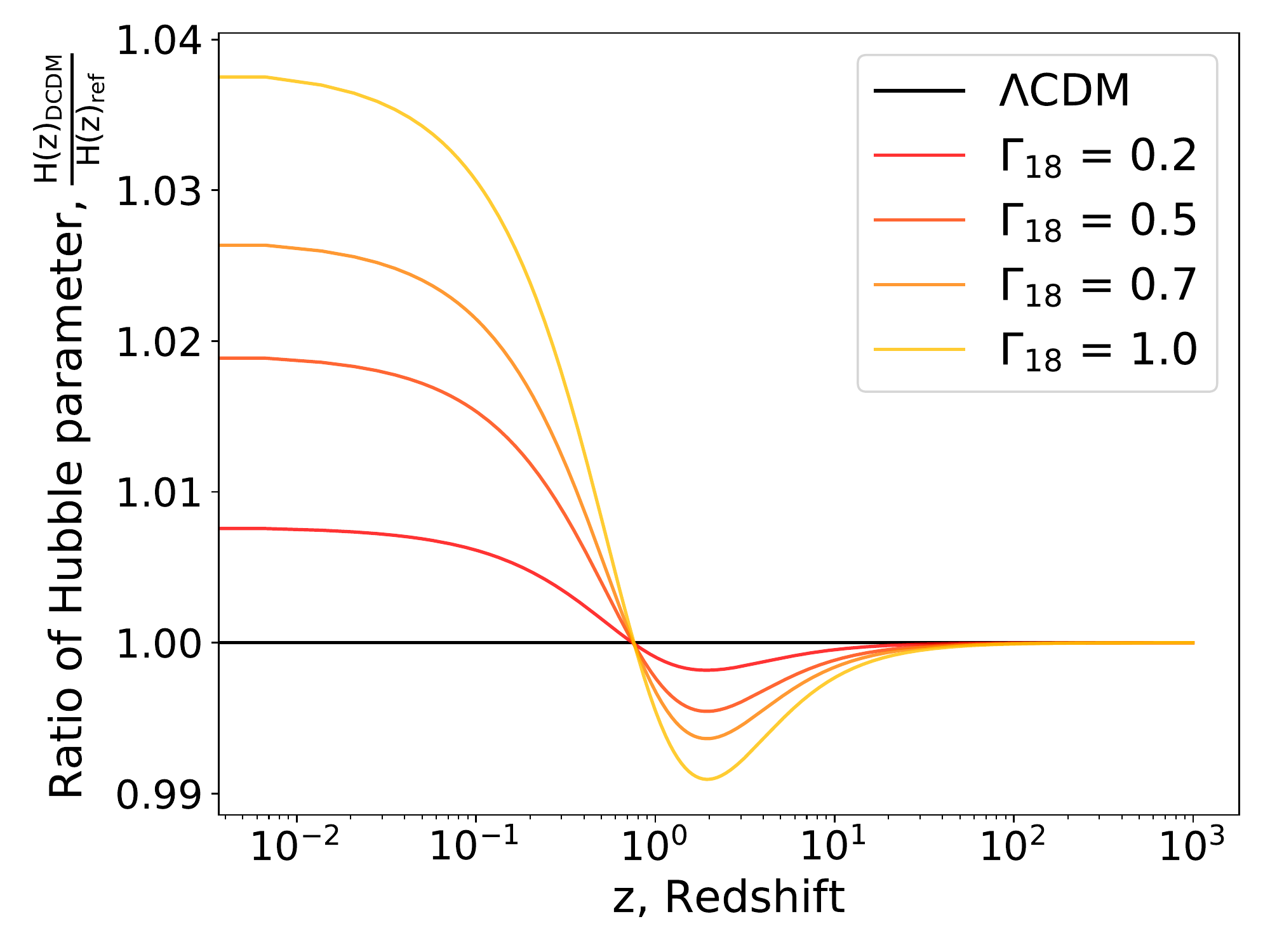}}
\end{minipage}%
\begin{minipage}{.5\linewidth}
\centering
\subfloat[]{\label{fig:cosmo_quant:omega}\includegraphics[scale=.35]{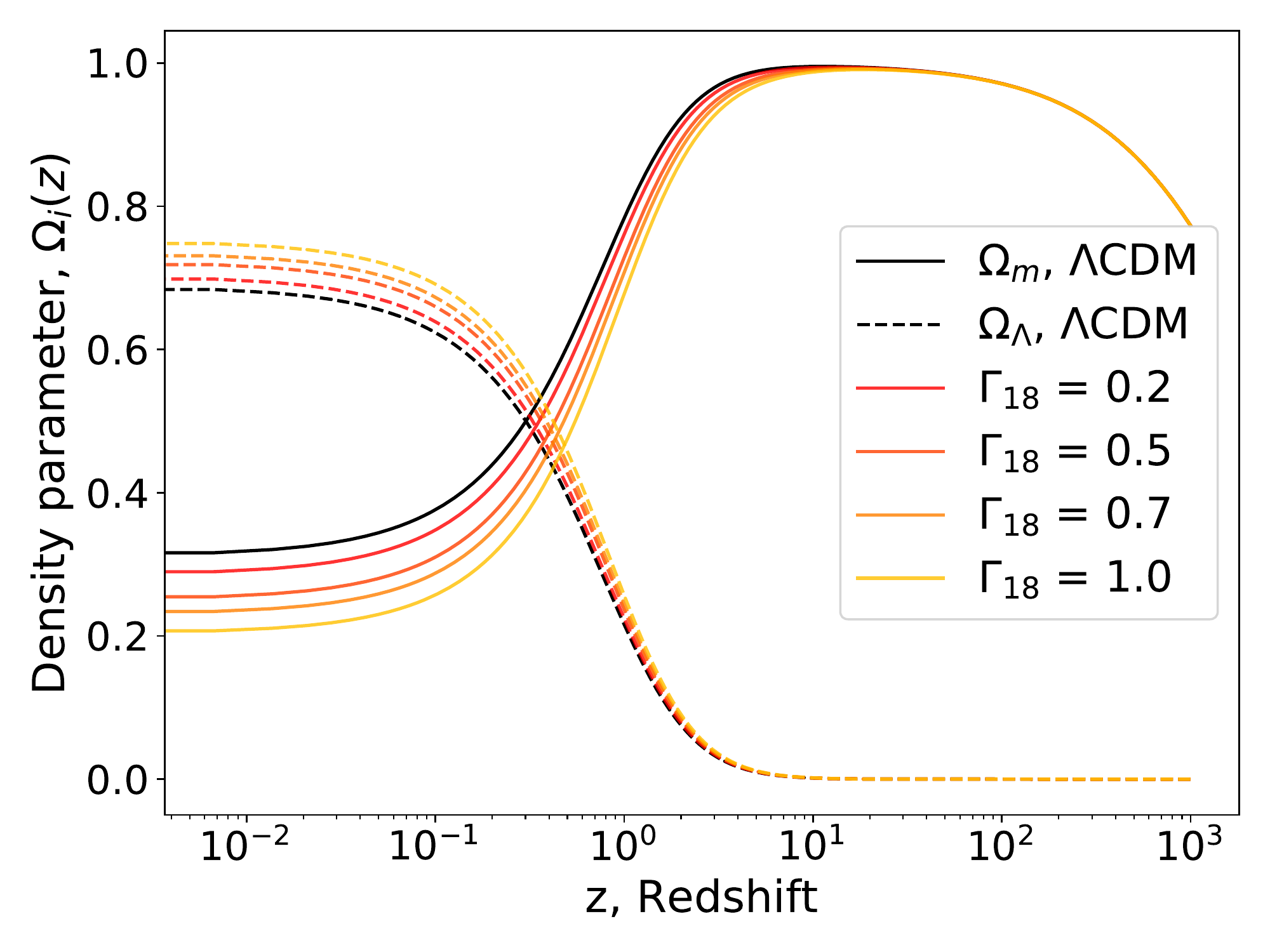}}
\end{minipage}\par\medskip
\caption{ \textit{Panel (a)}: Ratio of the Hubble parameter $H(z)$ for various values of the decay rate with respect to the corresponding quantity in the reference $\Lambda$CDM model, as a function of the redshift. Note that, for high values of decay rate, the Hubble parameter increases by a few percent (decreases) at later times (early times) with respect to the $\Lambda$CDM prediction.  \textit{Panel (b)}: Evolution of the matter (solid lines) and dark energy (dashed lines) density parameters with redshift. Note that for larger values of decay rate, the matter-$\Lambda$ equality (redshift at which solid and dashed curves overlap) occurs at higher redshift than in a universe with stable dark matter. }
\end{figure}

The effects of DCDM on background quantities is summarized in Figures \ref{fig:cosmo_quant:Hz} and \ref{fig:cosmo_quant:omega}, where we show the evolution of the Hubble parameter, the matter density parameter and the dark energy density parameter. In these plots, the angular size of sound horizon $\theta_\text{s}$ is kept fixed to a reference value. This choice is motivated by the fact that current CMB data constrain $\theta_\text{s}$ very precisely. In order to keep the angular size of sound horizon fixed while varying the dark matter decay rate, the dark energy density must vary accordingly. This modifies the evolution of the Hubble parameter, which is smaller (larger) than in $\Lambda$CDM at earlier (later) times. The epoch of matter-$\Lambda$ equality is also shifted towards earlier times.

\begin{figure}[tbp]
\centering
\includegraphics[width=0.80\textwidth,clip]{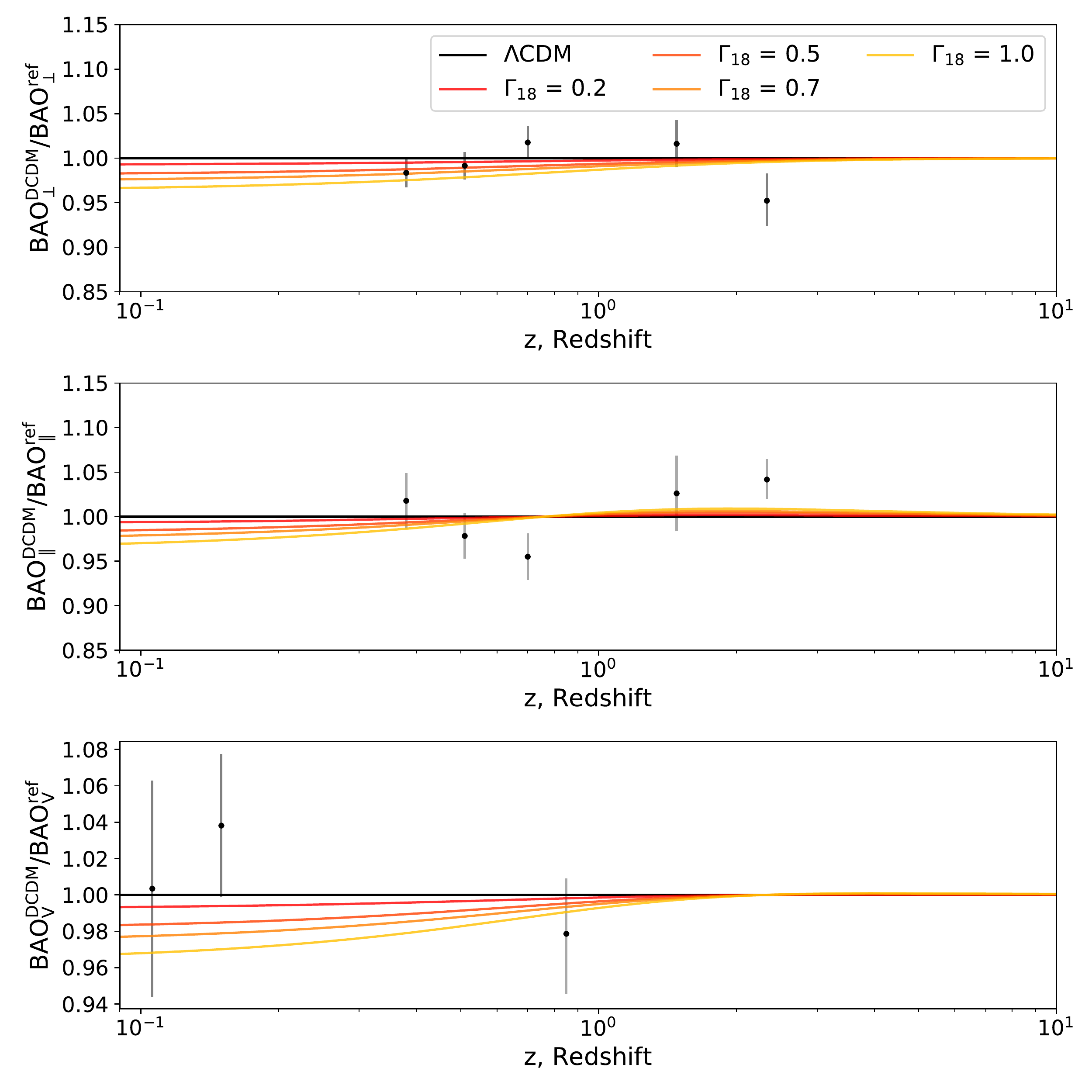}
\hfill
\caption{\label{fig:DM_DH_SDSS} Evolution of the angular BAO scales, given in terms of dimensionless quantities $D_M(z)/r_d$ (top panel), $D_H(z)/r_d$ (centre) and $D_V(z)/r_d$ (bottom), with redshift, normalized to the values of the same quantities in the case of stable DM. Top and middle panel data points are measurements from BOSS DR12 (galaxies) and eBOSS DR16 (galaxies, quasars, Lyman-$\alpha$), bottom panel measurements are from galaxies from 6dFGS, SDSS DR7 and eBOSS DR16. See Section \ref{sec:data_param_est} for details on the data. }
\end{figure}

Baryon acoustic oscillations (BAO) are powerful probes of the background dynamics. BAO are sensitive to the matter dynamics in the matter and dark energy dominated epochs. In order to put the BAO measurements in perspective in the DCDM scenario, in Figure \ref{fig:DM_DH_SDSS} we plot the ratio of the transverse BAO scale $D_M/r_d$ (top panel), longitudinal BAO scale $D_H/r_d$ (centre panel) and isotropic BAO scale $D_V/r_d$ (bottom panel) with respect to the standard $\Lambda$CDM case for the same decay rates as in the previous plots. In the relevant panels, we include measurements of $D_M(z)/r_d$, $D_H(z)/r_d$ and $D_V(z)/r_d$ as described in the caption (see Section \ref{sec:data_param_est} for details on the data). The behaviour of the curves can be easily explained with the evolution of the Hubble parameter described above.

\begin{figure}[tbp]
\centering
\includegraphics[width=1.0\textwidth,clip]{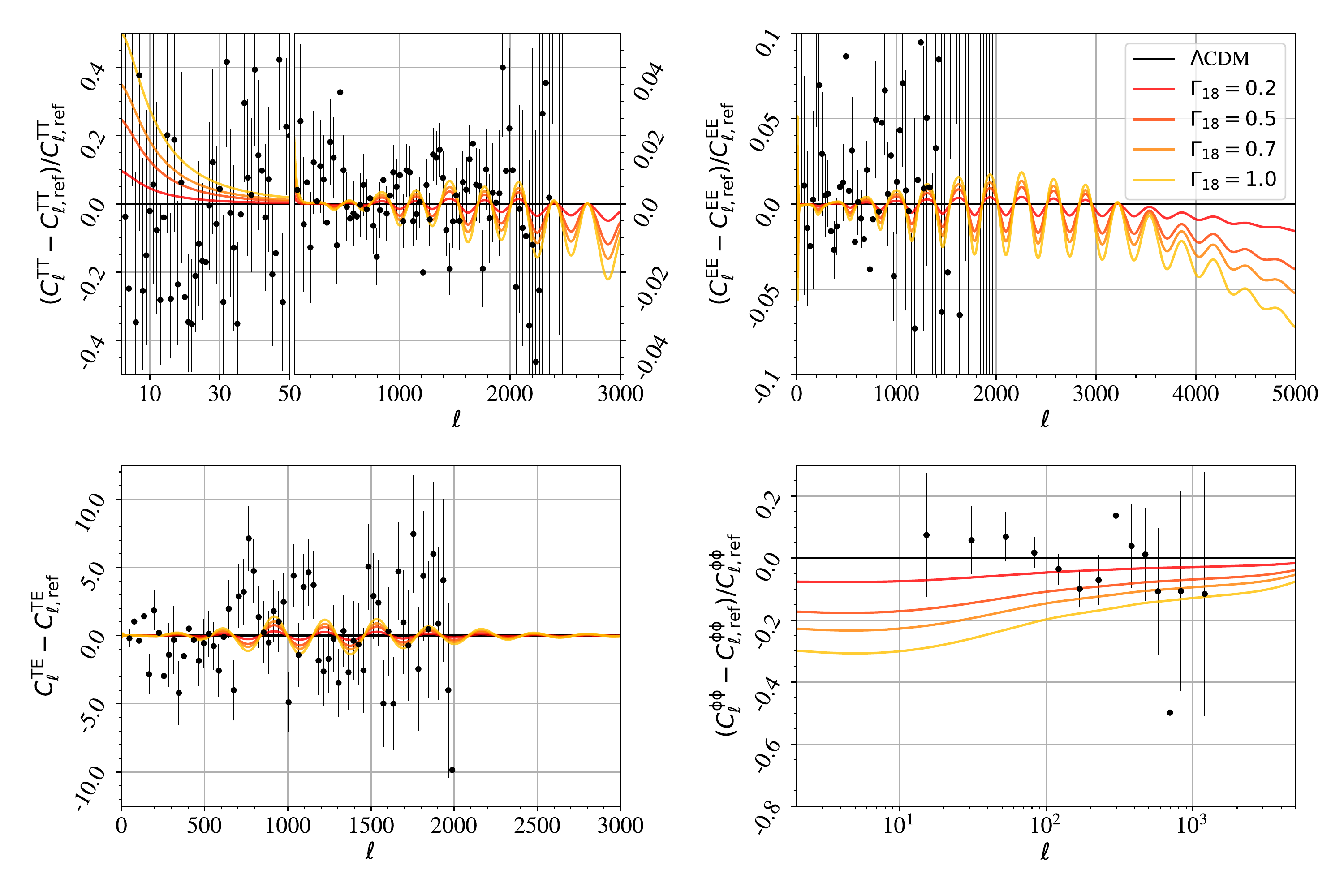}
\hfill
\caption{\label{fig:CMB_Decay} From the top left going clockwise, comparison of the CMB temperature (TT), polarization (EE), temperature-polarization (TE) and lensing potential ($\Phi\Phi$) power spectra for the standard case of stable dark matter ($\Lambda$CDM) with the DCDM scenario. In the presence of decaying dark matter, we have the following signatures: enhancement of the late-ISW effect (increase of power in the low-$\ell$ region of the TT spectrum); suppression of the gravitational lensing effect (decrease of amplitude of the $\Phi\Phi$ spectrum, reduced smoothing and increased damping of the high-$\ell$ region of the CMB primary anisotropies spectra).
}
\end{figure}

We now move to the effects of DCDM on cosmological perturbations. Figure~\ref{fig:CMB_Decay} shows the relative difference of CMB temperature (top left), polarization (top right), temperature-polarization (bottom left) and lensing potential (bottom right) power spectra with respect to the reference case of stable DM for different values of the DM decay rate. 

The longer duration of the $\Lambda$-dominated epoch in presence of DCDM is responsible for one of the main effects of DCDM on the CMB temperature power spectrum.
Indeed, the different evolution of gravitational potentials in the late Universe enhances the late-integrated Sachs Wolfe (late-ISW) effect. This is visible in the low-$\ell$ (large-scale) region of the CMB temperature power spectrum, where power is enhanced for higher decay rates. To better visualize the effect, in the top left panel of Figure~\ref{fig:CMB_Decay}, we blow up on the first 30 $\ell$ multipoles using a linear scale. Note that the y-axis on the right of the panel refers to the $\ell>30$ region, which is instead reported in logarithmic scale.
 
In this high-$\ell$ region of the temperature spectrum, DCDM manifests itself via the effect on CMB lensing. As the dark matter decay rate increases - thus depleting the abundance of dark matter at late times - the gravitational lensing effect due to the evolving large-scale structure is reduced. This is clearly visible in the damping of the lensing potential power spectrum (bottom right panel in Figure~\ref{fig:CMB_Decay}) and, in turn, in the oscillatory behaviour of the residuals in temperature (top left). 
More in detail, the zero-points of the oscillations remain the same, while the overall amplitude changes as the decay rate increases. When moving to higher values of $\Gamma_{18}$, the dark matter abundance at late times decreases. Hence, as stated above, the smearing effect on the CMB spectra due to lensing is reduced: the peaks are enhanced (and the relative difference with respect to the reference $\Lambda$CDM model is positive) and troughs are deeper (and the relative difference is negative). In addition, the reduced lensing effect manifests as an overall damping of power towards the smallest angular scales. This can be understood as follows: at the smallest scales, the unlensed spectra have very little power, and therefore benefit more of the transfer of power from large to small scales due to lensing. If the lensing effect is suppressed because of high decay rates, then there is less transfer of power with respect to the reference $\Lambda$CDM model.

The top right and bottom left panels in Figure~\ref{fig:CMB_Decay} show the effect of DCDM on the polarization spectra and its cross-correlation with temperature, respectively. 
First of all, we do not have the equivalent of the late-ISW in polarization. However, the effect of lensing on the polarization spectra is more pronounced compared to the temperature spectrum, as one could appreciate from the larger features at high multipoles, which have the same origin of those described in the case of the temperature spectra. Therefore, information from CMB polarization spectra are confirmed to be key \cite{Galli:2014kla}. We finally comment on the DCDM effects at large scales in polarization. In this work, we assume that the dark matter decays into dark radiation, thus not affecting directly the physics of reionization via the injection of high-energy photons. Therefore, we do not expect any degeneracy between the DCDM parameters and the reionization parameters. Nevertheless, as pointed out in~\cite{Poulin:2016nat}, DCDM could also affect the computation of (redshift-integrated) reionization parameters due to its effects on the late-time evolution of the universe. We do observe an oscillatory pattern at large scales in the residuals in polarization. However, the effect is well below cosmic variance - and thus unobservable - even for the exaggerated values of the decay rate that we assume to produce the figures in this work. 

\section{Results}
\label{sec:data_param_est}

We perform MCMC analyses to derive constraints on model parameters using the public version of \texttt{CLASS v2.9}\footnote{Get the latest \texttt{CLASS v3.2} at https://github.com/lesgourg/class\_public}~\cite{Blas:2011rf,Lesgourgues:2011re,Lesgourgues:2011rh} as Boltzmann solver and \texttt{MontePython v3.4}\footnote{The new \texttt{MontePython v3.6}, featuring the new eBOSS DR16 likelihoods used for this paper, will be available soon at https://github.com/brinckmann/montepython\_public}~\cite{Brinckmann:2018cvx,Audren:2012wb} as the MCMC engine. We sample, imposing flat priors, the following set of cosmological parameters: the baryon density parameter $\omega_b$, the present day density parameter of total dark matter if it were stable $\omega^\text{DCDM}_\text{ini}$ (see Equation \ref{eq:omega_dcdm}), the decay rate of unstable dark matter $\Gamma_{18}$, the angular size of the sound horizon at recombination $\theta_s$, the logarithmic amplitude $\log(A_s)$ at pivot scale $k_\mathrm{pivot}=0.05$ Mpc$^{-1}$ and the spectral index $n_s$ of the primordial scalar perturbations, and the optical depth to reionization $\taur$. Furthermore, we impose spatial flatness, we assume adiabatic initial conditions, and two massless and one massive neutrino of $0.06\,\eV$ with a standard contribution to the effective number of relativistic degrees of freedom of 3.046, consistent with the cosmology considered in Planck 2018 base analysis. We use the Gelman-Rubin criterion~\cite{Gelman:1992zz}, $R$, to assess the degree of convergence of our chains and have checked that the maximum value across the cosmological parameters is $R-1 \lesssim 0.01$ after removing burn-in and the non-Markovian part of the chains.

In the following, we consider two scenarios. We first assume that the dark matter is made by a single unstable component, and derive the corresponding constraints on the model parameters. We then relax this assumption and derive constraints assuming that only a fraction of the total dark matter abundance is unstable. The fraction of DCDM, $f_\text{DCDM}$, is defined as the ratio of unstable density ($\rho^\text{DCDM}_\text{ini}$) to total CDM density ($\rho^\text{CDM}_\text{ini} \equiv \rho_\text{CDM}+\rho^\text{DCDM}_\text{ini}$). Of course, the scenario in which the dark matter is completely composed of decayable specie corresponds to $f_\text{DCDM}=1$.\\
\\
In the analysis shown in this paper, we consider the following data:
\begin{itemize}
\item [(i)] CMB temperature and polarization angular power spectra as published in the Planck 2018 legacy release \cite{planck2016-l05} (referred to as "Planck T\&P" in the following),
\item [(ii)] CMB lensing reconstruction power spectrum from the Planck 2018 legacy release \cite{planck2016-l08} ("lensing" in the following),
\item [(iii)] Baryon Acoustic Oscillations (BAO) measurements from the Six-degree Field Galaxy Survey (6dFGS $z=0.106$)~\cite{Beutler:2011hx}, the Sloan Digital Sky Survey Data Release 7 (SDSS DR7) Main Galaxy Sample (MGS, $z_\mathrm{eff}=0.15$)~\cite{Ross:2014qpa,Howlett:2014opa}, the Baryon Oscillation Spectroscopic Survey (BOSS) DR12 galaxies ($z_\mathrm{eff}=0.38,0.51$)~\cite{BOSS:2016wmc}, as well as the Extended Baryon Oscillation Spectroscopic Survey (eBOSS) DR16~\cite{eBOSS:2020yzd} Luminous Red Galaxies (LRG, $z_\mathrm{eff}=0.7$)~\cite{Bautista:2020ahg,Gil-Marin:2020bct}, Emission Line Galaxies (ELG, $z_\mathrm{eff}=0.85$)~\cite{Tamone:2020qrl,deMattia:2020fkb}, quasars ($z_\mathrm{eff}=1.48$)~\cite{Hou:2020rse,Neveux:2020voa}, Lyman-$\alpha$
and Lyman-$\alpha\,-\,$quasar cross-correlation ($z_\mathrm{eff}=2.33$)~\cite{duMasdesBourboux:2020pck}. The combination of all of these BAO measurements are referred to as "BAO" in the following.
\item [(iv)] We add CMB measurements of B-mode polarization from the BICEP/Keck ground-based experiment (BK15) \cite{bib:BK15} dataset\footnote{ As we shall detail in the following, there is no degeneracy between the decay rate of dark matter and the tensor-to-scalar ratio. Thus, using the more recent BK18 dataset, currently not available in MontePython, would not significantly affect our results on $\Gamma_{18}$.} when we vary the tensor-to-scalar ratio.
\end{itemize}
We note that Lyman-$\alpha$ BAO measurements have been shown to be consistent with lower redshift BAO measurements by~\cite{Addison:2017fdm}, so we do not hesitate to combine the two. We also note that the analysis of BAO measurements are not expected to be problematic for the models considered in this paper, since the BAO peak is not shifted (see~\cite{Bernal:2020vbb} for an analysis of what kind of models might be affected).

\subsection{Results for $f_\text{DCDM}=1$}
We start our analysis assuming that all of dark matter in the Universe is unstable. The constraints on the dark matter decay rate are summarised in Figure~\ref{fig:gamma_constraints} (see Table~\ref{tab:all_param_derived} in the Appendix for the full constraints). In Figure~\ref{fig:gamma_ydcdm_2d}, we report the one- and two-dimensional posteriors on $\Gamma_{18}$ and a selection of other relevant cosmological parameters (see Figure ~\ref{fig:gamma_degen} for the full set of two-dimensional posteriors).

The Planck T\&P datasets alone put an upper limit on the DCDM decay rate of  $\Gamma_{18} < 0.175$ ($\tau_\text{DCDM} > 181$ Gyr) at 95\% C.L. This corresponds to a scenario in which the dark matter decays \textit{after} recombination. We note that Planck 2018 T\&P data alone put more stringent constraints on the dark matter decay rate than the limits reported using previous Planck data releases, see \cite{Audren:2014bca} for the 2013 release and \cite{Poulin:2016nat} for the 2015 release. This is due to different combined effects: i) use of the full-mission datasets in temperature and polarization; ii) improved reduction of polarization data with respect to the 2015 release; iii) improved constraints on the optical depth to reionization, and consequently on other cosmological parameters that are mostly degenerate with $\taur$.

For the scenario allowed by current data, i.e., dark matter decaying after recombination, we expect that the constraining power in the DCDM parameters comes from the observation of the late-ISW effect (albeit partly washed out by cosmic variance affecting the very large angular scales) and, mostly, from CMB lensing~\cite{Poulin:2016nat}. Indeed, the inclusion of small-scale CMB data marks almost an order-of-magnitude improvement of the constraints on $\Gamma_{18}$ in this work with respect to, e.g., the very first analyses using WMAP data~\cite{Lattanzi:2013uza}. 

When complemented by probes sensitive to low redshift evolution, i.e. lensing power spectrum and BAO measurements, the upper limit from Planck T\&P is further constrained at 95\% C.L. to $\Gamma_{18} < 0.136$ ($\tau_\text{DCDM} > 234$ Gyr) and $\Gamma_{18} < 0.129$ ($\tau_\text{DCDM} > 246$ Gyr), with lensing alone and with lensing plus BAO, respectively. We note that, while the combination of Planck T\&P and lensing almost saturates the constraining power on $\Gamma_{18}$, the inclusion of BAO is key to further improve the constraints on the remaining cosmological parameters, such as $\omega_\mathrm{ini}^\mathrm{DCDM}$.

\begin{figure}
\begin{center}
\centering
\small
\begin{tabular}{l|c|c}
\hline
Dataset\textbackslash Parameter & \multicolumn{1}{l|}{$\Gamma_{18} = \Gamma_\text{DCDM}/10^{-18} \text{s}^{-1}$} & \multicolumn{1}{l}{$\tau_\text{DCDM}$ Gyr} \\ \hline\hline
Planck T\&P       & $<0.175$                                                                & $>181$                             \\ \hline
+ lensing         & $<0.136$                                                                & $>234$                             \\ \hline
+ BAO               & $<0.129$                                                                & $>246$                             \\ \hline
\end{tabular}\\
\begin{minipage}[t]{.60\linewidth}
\centering
\includegraphics[width=\linewidth]{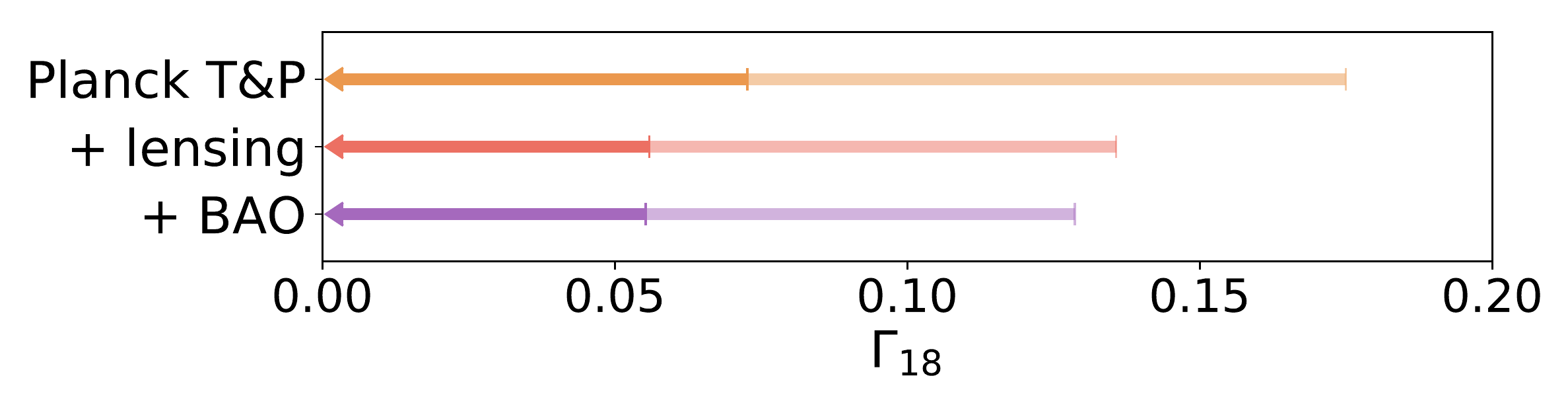}
\end{minipage}
\caption{The figure shows the 68\% (darker) and 95\% (lighter) C.L. bounds on dark matter decay rate $\Gamma_{18}$ and lifetime $\taudm$ for various datasets considered in this work: Planck 2018 CMB temperature and polarization spectra alone, with the Planck 2018 CMB lensing power spectrum, and additionally with BAO measurements from eBOSS DR16 and earlier datasets (see Section~\ref{sec:data_param_est} for more details). In the accompanying table we report the 95\% C.L. bounds.
}
\label{fig:gamma_constraints}
\end{center}
\end{figure}

\begin{figure}
\centering
\includegraphics[scale=.55]{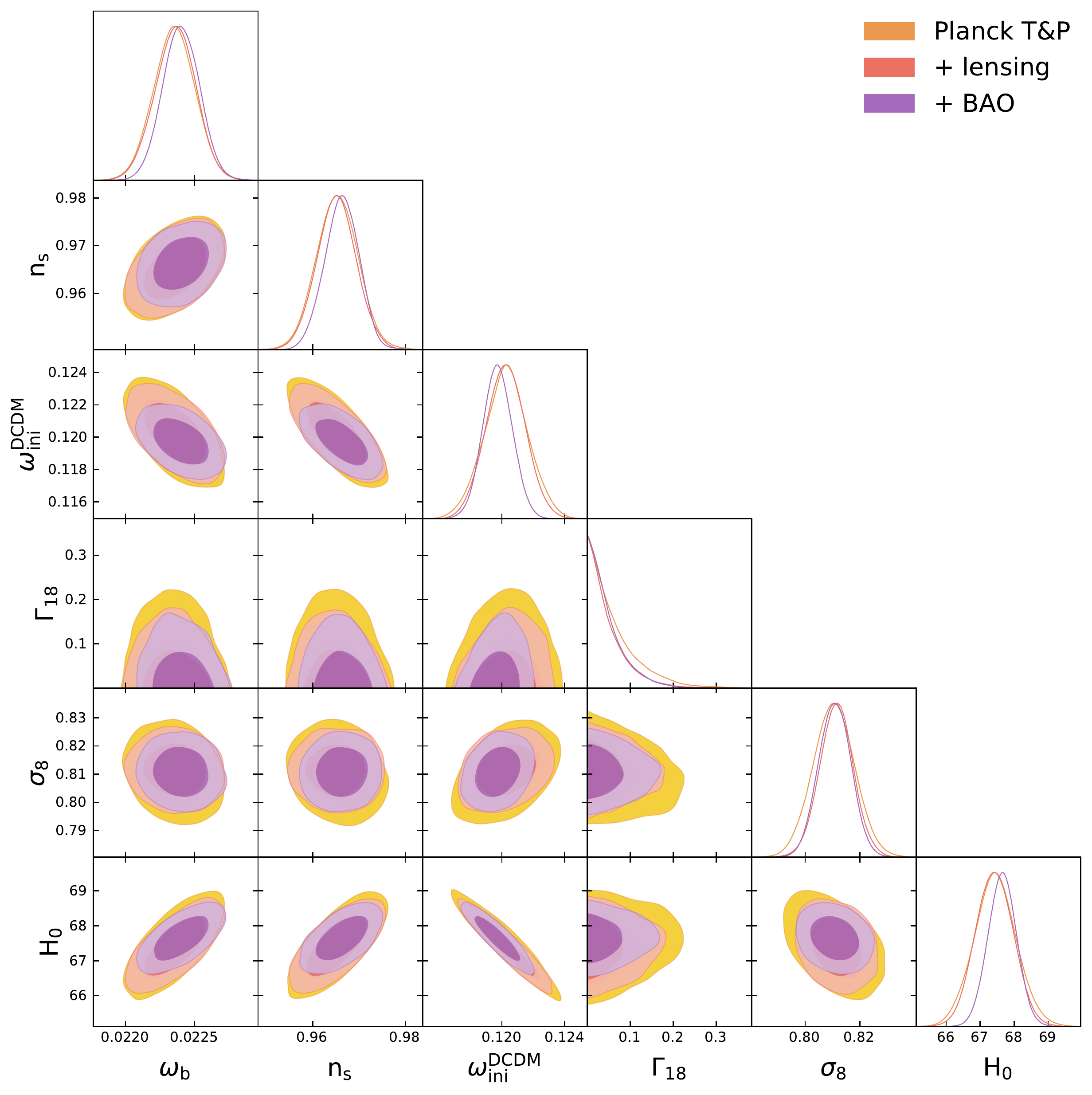}

\caption{\label{fig:gamma_ydcdm_2d} 2D contours between a number of standard cosmological parameters and the primordial dark matter density $\omega_\text{ini}^\text{DCDM}$ and dark matter decay rate $\Gamma_{18}$. The contours show the relative constraining power of the various probes considered in this work: Planck 2018 CMB temperature and polarization spectra alone (grey), in combination with the Planck 2018 CMB lensing power spectrum (red), and additionally with BAO measurements (blue) from eBOSS DR16 and earlier datasets (see Section~\ref{sec:data_param_est} for more details). The darker (lighter) regions are the 68\% (95\%) confidence level contours.
}
\end{figure}

We conclude this section with a brief discussion on major degeneracies between the DCDM decay rate $\Gamma_{18}$ and the remaining cosmological parameters. In the scenario allowed by current data (decay after recombination), we expect a major degeneracy between the decay rate $\Gamma_{18}$ and the dark energy parameter $\Omega_\Lambda$, as both affect the amplitude of the late-ISW effect. Indeed, as shown in Figure~\ref{fig:gamma_ydcdm_2d}, the parameter $\Gamma_{18}$ is almost uncorrelated with the other sampled parameters. The mild correlation between $\Gamma_{18}$ and $\omega_\text{ini}^\text{DCDM}$ is understandable, as a greater decay rate of DCDM should come with a slightly larger dark matter density at early times, so a portion of dark matter is allowed to decay.

As expected, adding lensing and BAO data to Planck T\&P data improves the constraint on $\Gamma_{18}$, since they both provide a late time estimation of the dark matter density. Further details are reported in Appendix~\ref{sec:appA}, where we show the correlation matrix for the full set of sampled paremeters and a subset of derived cosmological parameters. The full set of cosmological constraints are reported in Table~\ref{tab:all_param_derived}.

\subsection{Results for $f_\text{DCDM}\ne1$}
In our base analysis, we assumed that all of dark matter is unstable. This assumption is rather ad-hoc, as there is no reason to believe that dark matter is composed of only one species of particles. In this subsection, we consider an extension of our base analysis where we assume that only a fraction of the total dark matter is allowed to decay. We proceed as follows: we fix the fraction of unstable dark matter to a given value and derive constraints on the model parameters in this scenario. This allows us to obtain the upper bound on the DM decay rate as a function of the unstable dark matter fraction. 

\begin{figure}
\centering
\includegraphics[scale=0.5]{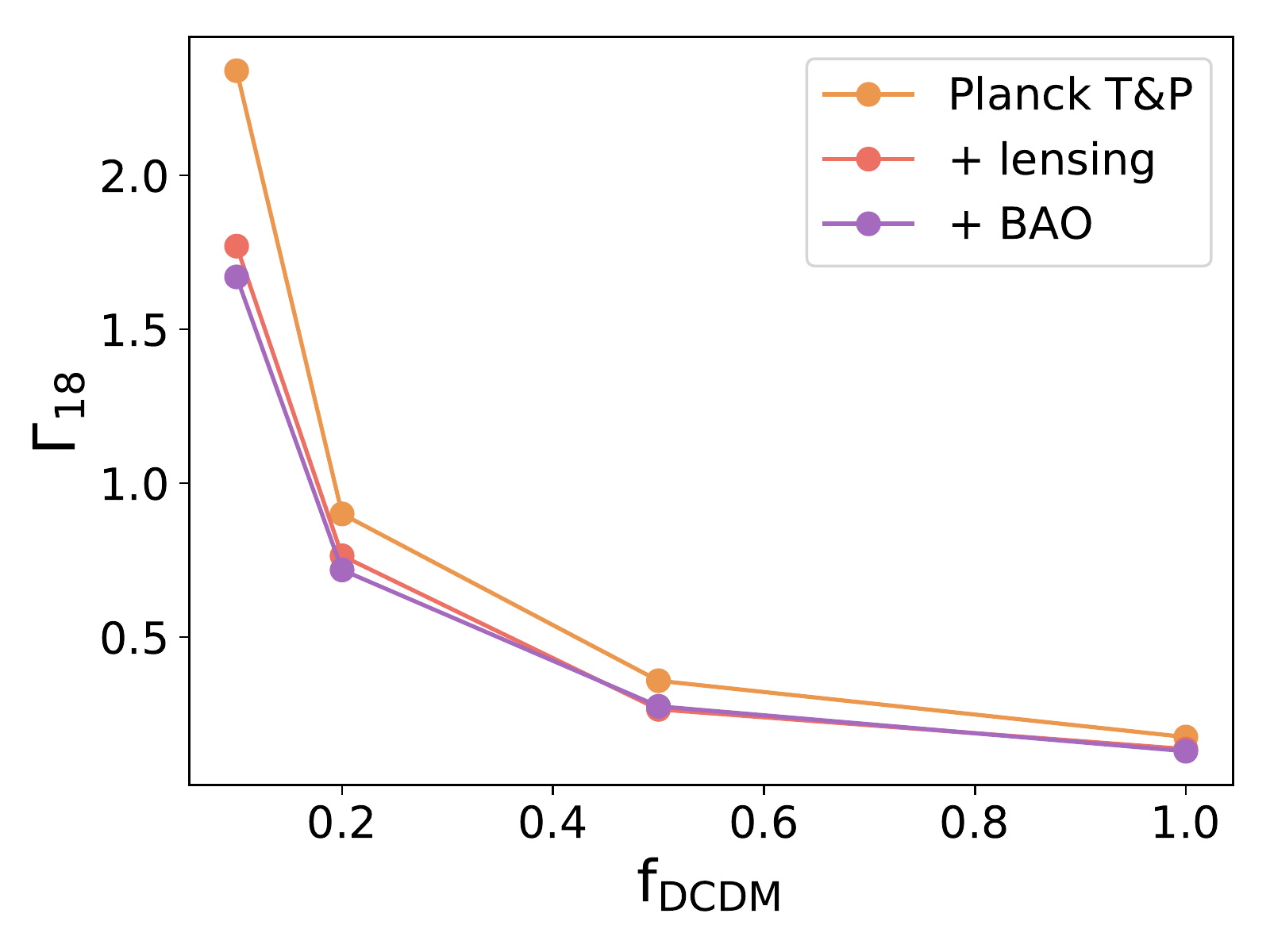}
\caption{The 95\% C.L. upper limit of decay rate for various fixed values of DCDM fraction for the various datasets considered in this work. As explained, for larger values of fraction, the upper limit on the decay rate is constrained to smaller values while for smaller values of the fraction, the limit is relaxed and DCDM can be allowed to decay quickly, even depleting before matter-radiation equality.
}
\label{f_DCDM_gamma}
\end{figure}

In Figure \ref{f_DCDM_gamma}, we show the 95\% confidence limit on decay rate as a function of the DCDM fraction for the Planck T\&P (blues), T\&P+lensing (orange), T\&P+lensing+BAO (green) datasets. In Table~\ref{tab:frac_neq_decay}, we report the 95\% C.L. limits on the decay rate for the different values of $f_\mathrm{DCDM}$ considered in this work. It can be seen that with a small decay rate (long lived dark matter) $f_\text{DCDM}$ can be a large number (a major fraction of the total CDM can be unstable). On the other hand, for a small fraction of unstable dark matter, the decay rate can be very large yet in agreement with the data. Of course, the case where the fraction is one corresponds to our baseline case discussed in the previous subsection.
As one can see from Table~\ref{tab:frac_neq_decay}, if only a fraction of the dark matter is unstable and the decay rate is small compared to the age of the Universe ($\Gamma_\mathrm{DCDM} \lesssim H_0$), cosmological data effectively constrain the combination $\Gamma_\mathrm{DCDM} f_\text{DCDM}$ (also noted in Ref.~\cite{Poulin:2016nat}). As an example, we see that the bound on the decay rate can be relaxed by a factor of ten if only 10\% of the dark matter is unstable. 

\begin{table}[]
\begin{tabular}{c|c|c|c}
\hline
Dataset                      & Model\textbackslash Parameter       & $\Gamma_{18} = \Gamma_\text{DCDM}/10^{-18} \text{s}^{-1}$ & $\tau_\text{DCDM}$ {[}Gyr{]} \\ \hline\hline
\multirow{4}{*}{Planck T\&P} & $f_\text{DCDM} = 1.0$ & $<0.175$                                           & $>181$              \\ \cline{2-4} 
                             & $f_\text{DCDM} = 0.5$ & $<0.358$                                           & $>88.4$             \\ \cline{2-4} 
                             & $f_\text{DCDM} = 0.2$ & $<0.900$                                           & $>35.2$             \\ \cline{2-4} 
                             & $f_\text{DCDM} = 0.1$ & $<2.34$                                            & $>13.6$             \\ \hline\hline
\multirow{4}{*}{+ lensing}   & $f_\text{DCDM} = 1.0$ & $<0.136$                                           & $>234$               \\ \cline{2-4} 
                             & $f_\text{DCDM} = 0.5$ & $<0.265$                                           & $>119$              \\ \cline{2-4} 
                             & $f_\text{DCDM} = 0.2$ & $<0.764$                                           & $>41.5$             \\ \cline{2-4} 
                             & $f_\text{DCDM} = 0.1$ & $<1.77$                                            & $>17.9$             \\ \hline\hline
\multirow{4}{*}{+ BAO}       & $f_\text{DCDM} = 1.0$ & $<0.129$                                           & $>246$             \\ \cline{2-4} 
                             & $f_\text{DCDM} = 0.5$ & $<0.275$                                           & $>115$              \\ \cline{2-4} 
                             & $f_\text{DCDM} = 0.2$ & $<0.718$                                           & $>44.2$             \\ \cline{2-4} 
                             & $f_\text{DCDM} = 0.1$ & $<1.67$                                            & $>18.9$             \\ \hline
\end{tabular}
\caption{We show the 95\% C.L. bounds on the dark matter decay rate $\Gamma_\mathrm{DCDM}$ and lifetime $\tau_\mathrm{DCDM}$ for the different cases of $f_\text{DCDM}$ (fraction of dark matter that is unstable) and different dataset combinations considered in this work: Planck 2018 CMB temperature and polarization spectra alone, with Planck 2018 CMB lensing power spectrum, and additionally with BAO measurements from eBOSS DR16 and earlier datasets (see Section~\ref{sec:data_param_est} for more details).}
\label{tab:frac_neq_decay}
\end{table}

\subsection{One-parameter extensions of the DCDM model}\label{sec:ext_models}
 In this section, we elaborate on some one-parameter extensions to the DCDM model. In particular, we explore models in which the sum of the neutrino masses, $\Sigma m_\nu$, the curvature parameter, $\Omega_k$, and the tensor-to-scalar ratio, $r$, are allowed to vary, respectively, together with the other parameters of the baseline DCDM model. We consider this set of one-parameter extensions since the additional parameters may affect cosmological probes in a way that could compensate for the effects of decaying dark matter. When considering extended models, we only explore the case where all of dark matter is unstable, i.e., $f_\text{DCDM}=1$.

In \cite{Poulin:2016nat}, the authors explored the degeneracy between the DCDM decay rate and neutrino mass. The neutrinos behave like radiation at early cosmological times, while later, after the non-relativistic transition of neutrinos, they evolve like a matter component. We would expect the opposite trend for DCDM evolution: before a significant amount of DCDM decays into dark radiation, the fluid behaves like a matter component. However, at late times it also contributes to the radiation content of the universe through the decay products. Therefore, we could expect that the effect of neutrinos could be cancelled by a suitable combination of decay rate and DCDM fraction. Ref. \cite{Poulin:2016nat} reports that the small-scale CMB temperature and polarization measurements effectively break the degeneracy between these parameters as they affect the spectra in different ways. They, therefore, obtain independent constraints on the two parameters. We reassess the degeneracy with our combined datasets (Planck T\&P, lensing and BAO) by varying the total neutrino mass, keeping the same effective number of relativistic degrees of freedom, $N_\mathrm{eff}= 3.046$, and obtain results consistent with \cite{Poulin:2016nat}, as shown in Figure \ref{fig:muLDDM+LCDM} of the appendix.

The curvature density parameter, $\Omega_k$, like DCDM, affects the late-ISW part of the CMB temperature spectra. The Planck 2018 datasets \cite{bib:Planck2018} show a preference for a universe with a negative curvature, but the addition of the BAO datasets makes the estimates consistent with a flat $\Lambda$CDM model, as demonstrated in \cite{eBOSS:2020yzd}. Therefore, the BAO datasets are essential in establishing flatness in the $\Lambda$CDM model. In \cite{Poulin:2016nat}, the degeneracy between the curvature density parameter and the decay rate was explored and no degeneracy was found. We reassess this with our combined datasets and, similar to \cite{Poulin:2016nat}, we find a curvature density parameter consistent with a flat $\Lambda$CDM model, as in \cite{eBOSS:2020yzd}. We show the triangle plot in Figure \ref{fig:OkLDDM+LCDM} in the appendix.

In Figure \ref{fig:rLDDM+LCDM}, we show the posteriors for the DCDM + $r$ model and compare with the constraints on the tensor-to-scalar ratio, $r$, and $\Gamma_{18}$ in $\Lambda$CDM + $r$ or DCDM, respectively. The 1D posteriors of $\Gamma_{18}$ are nearly identical between DCDM and DCDM + $r$ models, suggesting that there is no degeneracy between these two parameters. 

In Table \ref{tab:ext_models}, we report constraints for the parameters of the extended models and $\Gamma_{18}$, comparing them with those obtained in the baseline DCDM model and in the standard $\Lambda$CDM model. We note that the constraints on the decay rate degrade slightly in the extended models, but they are generally consistent with our baseline constraints. The same is also true for the additional parameters when compared with their values in the minimal extensions to $\Lambda$CDM (with stable dark matter).

\begin{table}[h]
\centering
{\fontsize{10pt}{18pt}\selectfont
\begin{tabular}{l|c|c|c|c|c|c}
\hline
\multicolumn{1}{l|}{\textbf{\begin{tabular}[l]{@{}l@{}}Parameter/ \\ Model\end{tabular}}} & \multicolumn{1}{c|}{\begin{tabular}[c]{@{}c@{}}$\Gamma_{18}$\\ {\fontsize{8pt}{18pt}$\Gamma_\text{DCDM}/10^{-18} \text{s}^{-1}$}\end{tabular}} & \multicolumn{1}{c|}{\begin{tabular}[c]{@{}c@{}}$\sum m_\nu$\\ {\fontsize{8pt}{18pt}$\eV$}\end{tabular}} & \multicolumn{1}{c|}{$\Omega_{k} \cdot10^{-3}$} & \multicolumn{1}{c|}{$r$} & \multicolumn{1}{c|}{$\sigma_8$} & \multicolumn{1}{c}{\begin{tabular}[c]{@{}c@{}}H$_0$\\ {\fontsize{8pt}{18pt}kms$^{-1}$Mpc$^{-1}$}\end{tabular}} \\ \hline\hline
$\Lambda$CDM  & $\equiv$ 0.0 & $\equiv$ 0.06 & $\equiv$ 0.0 & $\equiv$ 0.0 & $0.8112_{- 0.0062}^{+ 0.0060}$ & $67.63 \pm 0.42$ \\ \hline
DCDM  & $<0.129$ & $\equiv$ 0.06 & $\equiv$ 0.0 & $\equiv$ 0.0 & $0.8110_{- 0.0062}^{+ 0.0056}$ & $67.67 _{- 0.44}^{+ 0.39}$ \\ \hline
\hline
$\Lambda$CDM+$\sum m_\nu$  & $\equiv$ 0.0 & $<0.132$ & $\equiv$ 0.0 & $\equiv$ 0.0 & $0.8129 _{- 0.0079}^{+ 0.0126}$ & $67.71 _{- 0.50}^{+ 0.55}$ \\ \hline
DCDM+$\sum m_\nu$  & $<0.141$ & $<0.128$ & $\equiv$ 0.0 & $\equiv$ 0.0 & $0.8128 _{- 0.0094}^{+ 0.0129}$ & $67.75 _{- 0.52}^{+  0.59}$\\ \hline\hline
$\Lambda$CDM+$\Omega_k$ & $\equiv$ 0.0 & $\equiv$ 0.06 & $0.1_{-2.0}^{+ 1.8}$ & $\equiv$ 0.0 & $0.8116 _{- 0.0072}^{+ 0.0074}$ & $67.64 _{-0.64}^{+ 0.62}$ \\ \hline
DCDM+$\Omega_k$  & $<0.147$ & $\equiv$ 0.06 & $-0.1_{-1.9}^{+ 2.0}$ & $\equiv$ 0.0 & $0.8106 _{- 0.0074}^{+ 0.0073}$ & $67.63 _{- 0.65}^{+ 0.63}$ \\ \hline\hline
$\Lambda$CDM+r & $\equiv$ 0.0 & $\equiv$ 0.06 & $\equiv$ 0.0 & $<0.065$ & $0.8119_{- 0.0063}^{+ 0.0058}$ & $67.58 \pm 0.41$\\ \hline
DCDM+r & $<0.141$ & $\equiv$ 0.06 & $\equiv$ 0.0 & $<0.066$ & $0.8116_{- 0.0069}^{+ 0.0062}$ & $67.64_{- 0.44}^{+ 0.43}$\\ \hline \end{tabular}}
\caption{We summarize the constraints on $\Gamma_{18}$ in our baseline model along with those obtained in the extended models. As explained in \ref{sec:ext_models}, we use our baseline combined datasets to constrain these extended models except in the case when we vary the tensor-to-scalar ratio, in which case we also include the BK15 dataset. While the constraints on $\Gamma_{18}$ are slightly weaker in the extended models, they are generally consistent with our baseline DCDM model. We also show the constraints on the additional parameters along with their values in the standard scenario of stable dark matter. We point out that even in the extended models, the values of $\sigma_8$ and H$_0$ remain essentially consistent with their values in the $\Lambda$CDM universe, hence the $\sigma_8$ and H$_0$ tensions persist in these models. All numbers quoted with two-sided bounds are 68\% confidence intervals while those with one-sided bounds are the 95\% upper limits. Parameters which are kept fixed in a given model are represented with an $\equiv$ sign.}
\label{tab:ext_models}
\end{table}

\section{Implications for particle physics models}\label{sec:particle_physics}

We consider the Majoron as an example of a dark matter candidate that mostly decays into invisible particles, in particular neutrinos.
It has been proposed that the spontaneous breaking of ungauged lepton number might be at the origin of the small neutrino masses.
If that is the case, the Majoron $J$ is the (massless) Nambu-Goldstone boson arising from such a breaking \cite{Chikashige:1980ui,Gelmini:1980re,Schechter:1981cv}. The fundamental parameter of the model
is the scale of lepton number-breaking $\fL$. This (high) energy scale controls the couplings of the Majoron to neutrinos and thus the neutrino masses. The Majoron is naturally long-lived, since its couplings are suppressed by the scale $\fL$.
In the following, we will assume the so-called see-saw limit $v \ll \fL$, with $v=246\,\mathrm{GeV}$ being the vacuum expectation value of the Higgs boson.

If, additionally, lepton number is also explicitly broken, the Majoron might acquire a mass $m_J$, making it a \emph{pseudo} Nambu-Goldstone boson. This might be caused by nonpertubative gravitational effects \cite{Rothstein:1992rh, Akhmedov:1992hi, Alonso:2017avz}, or by terms in the Lagrangian that softly violate lepton number \cite{Gu:2010ys}. Such a massive Majoron would be a suitable DM candidate \cite{Rothstein:1992rh,Berezinsky:1993fm,Lattanzi:2007ux,Bazzocchi:2008fh,Lattanzi:2013uza,Queiroz:2014yna,Rojas:2017sih,Garcia-Cely:2017oco,Brune:2018sab}

In the minimal singlet Majoron model, the only tree level couplings of the Majoron are to neutrinos and possibly to the Higgs boson.
In particular, in the see-saw limit the Lagrangian describing the interaction between the Majoron and the light active neutrinos $\nu_j$ with masses\footnote{To be more precise, the $\nu_j$'s appearing in the equation are the light mass eigenstates. Even though these are in general a superposition of active and sterile (singlet) neutrinos, in the seesaw limit the mixing is dominated by the active states.} $m_j$ ($j=1,2,3$) at low energies is
\begin{equation}
    {\mathcal L_{J\nu\nu}} = \frac{i J}{2\fL} \sum_{i=1}^{3} m_i \bar\nu_i \gamma_5 \bar \nu_i \, .
\end{equation}
In the limit $m_J \gg m_i$, this induces the Majoron decay into two neutrinos with a rate:
\begin{equation}
\Gamma_{J\to\nu\nu} = \frac{m_J}{16 \pi \fL^2} \sum_{i=1}^3 m_i^2 \, .
\end{equation}

The sum of neutrino masses squared that appear on the right-hand side of the equation is experimentally well constrained. In particular, flavour oscillation measurements of the mass-squared differences imply that $\sum m_i^2 > 2.5 \times 10^{-3}\,\eV^2 $ ($ > 4.9 \times 10^{-3}\,\eV^2 $) for normal (inverted) mass ordering. An upper limit might be derived by further considering cosmological bounds on the neutrino mass scale. However, for the purpose of obtaining a constraint on $\fL$, it is enough to conservatively fix $\sum m_i^2$ to its lower bound from oscillations. The bound $\Gamma_{J\to\nu\nu} < 0.129 \times 10^{-18} \,\mathrm{s}^{-1} = 8.5 \times 10^{-35}\,\eV$ from Planck T\&P plus CMB lensing and BAO data thus conservatively implies:
\begin{equation}
    \frac{\fL}{\mathrm{GeV}} > \left\{ \begin{array}{l}
     2.4\times 10^7 \displaystyle \left(\frac{m_J}{\mathrm{keV}}\right)^{1/2} \quad\textrm{(normal ordering)}\,  , \\[0.4cm]
     3.4\times 10^7 \displaystyle \left(\frac{m_J}{\mathrm{keV}}\right)^{1/2} \quad\textrm{(inverted ordering)}\, ,
    \end{array} \right.
\end{equation}
if the Majoron makes 100\% of the dark matter. 

\section{Discussion and Conclusions}
\label{sec:conclusions}
In this work, we shed light on the stability of dark matter focusing on the decay into massless particles. 
We use the latest Planck 2018 datasets (temperature and polarization, T\&P, and lensing) together with BAO measurements from eBOSS DR16 and earlier datasets (see Section~\ref{sec:data_param_est} for more details) to constrain the lifetime of dark matter. We provide a measure of the relative constraining power between Planck T\&P, Planck T\&P plus CMB lensing, and T\&P plus CMB lensing and BAO data. 
Assuming that all the dark matter is unstable, the Planck T\&P datasets constrain the upper limit of the dark matter decay rate to $\Gamma_{18} = \Gammadm/10^{-18}\,\mathrm{s}^{-1}<0.175$ (lower limit of dark matter lifetime to $\taudm > 181$ Gyr) at 95\% C.L. 

The addition of CMB lensing improves the lower limit on DCDM lifetime (95\% limit) by $\approx 22$\% ($\Gamma_{18} <0.136$ and $\taudm>234$ Gyr). Further adding the BAO data to Planck T\&P plus lensing marginally improves the 95\% lower limit by $\approx$ 5\% ($\Gamma_{18} <0.129$ and $\taudm>246$ Gyr) over Planck T\&P plus lensing combination.

We also study the possibility that only a fraction $f_\mathrm{DCDM}$ of the total dark matter abundance is unstable. In agreement with previous studies~\cite{Poulin:2016nat}, we find that, for small-enough decay rates ($\Gamma_\mathrm{DCDM}\lesssim H_0$), the bounds on the decay rate can be relaxed with respect to the $f_\mathrm{DCDM}=1$ in such a way that the product $(\Gamma f)_\mathrm{DCDM}$ stays almost constant.

The results presented in this work have interesting implications for particle physics models. In particular, if we identify the Majoron as an example of dark matter candidate, the bounds we obtain on the decay rate, combined with a bound on the sum of the neutrino masses from flavour oscillation experiments, can be translated into a bound on the coupling of the Majoron to neutrinos, $f_L>2.4(3.4)\times 10^7\,(m_J/\mathrm{keV})^{1/2}\,\mathrm{GeV}$ assuming normal (inverted) neutrino mass ordering.

During the preparation of this manuscript, the authors of \cite{Simon:2022ftd} investigated the scenario of dark matter to dark radiation decay, for the scenarios of all dark matter decaying and only a fraction of dark matter decaying. 
They include in their analysis CMB temperature, polarization, and lensing power spectra from Planck 2018 in combination with full shape galaxy power spectrum analysis of BOSS DR12 data using EFTofLSS~\cite{DAmico:2020kxu,Chudaykin:2020aoj}, including mildly non-linear scales. This was the first time EFTofLSS had been used for a decaying dark matter analysis. They additionally include Type-Ia supernova data from Pantheon~\cite{Pan-STARRS1:2017jku} in their analysis. In addition to the EFTofLSS analysis, they do a complementary analysis using BAO data from 6dFGS, SDSS DR7, BOSS DR12 galaxies, including redshift-space distortion measurements ($f\sigma_8$) for the latter, as well as Lyman-$\alpha$ measurements from eBOSS DR14.
Our analysis differs from theirs in that we focus on conservative linear bounds and use newer BAO data from eBOSS DR16, which updated measurements from LRGs compared to BOSS DR12, added measurements for ELGs, and added measurements from quasars, alone and cross-correlated with Lyman-$\alpha$ measurements (see Section~\ref{sec:data_param_est} for a description of the data). To this end we developed the likelihood module in MontePython to include the latest eBOSS DR16 datasets reported in \cite{eBOSS:2020yzd}.\footnote{This will be made publicly available soon at https://github.com/brinckmann/montepython\_public with the new \texttt{MontePython v3.6}.} To be extra conservative, we omit supernova data over possible concerns of systematics and omit $f\sigma_8$ measurements over possible concerns that they were produced assuming $\Lambda$CDM. Despite these conservative choices, we find bounds on the lifetime of decaying dark matter $\taudm > 246$ Gyr (95\% C.L.), in general agreement with their bounds from Planck 2018 plus EFTofLSS and Pantheon of $\taudm > 249.6$ Gyr, as well as their Planck 2018 plus BAO, $f\sigma_8$ and Pantheon of $\taudm > 250.0$. Although additional data somewhat improves the bounds, this shows that we can already derive very stringent bounds on the dark matter lifetime when using only a very conservative selection of datasets. We note that Ref.~\cite{Nygaard:2020sow} also considered linear constraints from Planck 2018 plus BAO data from BOSS DR12 galaxies and we find general agreement with their results, although differences in analysis settings make a direct comparison of bounds on $\taudm$ only approximate, as even their "very long-lived" case does not correspond to $\fdm=1$ of our baseline case.

There are well-known inconsistencies within the standard $\Lambda$CDM model, namely the disagreement in $\sigma_8$ \cite{bib:HSCHamana:2019etx,bib:KiDSHeymans:2020gsg,DES:2021wwk} and H$_0$ \cite{Verde:2019ivm,DiValentino:2021izs,Schoneberg:2021qvd} estimated from different probes. While the possibility remains that these inconsistencies might simply be due to systematic effects, they might yet be indicative of new physics beyond the standard cosmological model. However, our analysis indicates that it seems unlikely that these tensions could be resolved exclusively by adding cold dark matter decay to very light particles, as we detail in the following. In the triangle plot of Figure \ref{fig:gamma_ydcdm_2d}, we show the degeneracy between $\Gamma_{18}$, H$_0$ and $\sigma_8$ and in Figure \ref{fig:corr_matrix} of the appendix we show more explicitly the correlation between these parameters. We see that an appreciable correlation exists when we consider only the Planck T\&P datasets. However, the strong constraints on the decay rate from the Planck T\&P datasets indicates "nearly" stable dark matter in the universe and, therefore, even with appreciable correlation between $\Gamma_{18}$, $\text{H}_0$, and $\sigma_8$, we obtain values consistent with the Planck 2018 baseline results, as shown in Table \ref{tab:all_param_derived}. When we further complement the Planck T\&P datasets with probes sensitive to the late-time dynamics of the universe, the degeneracy between the decay rate of DCDM with $\sigma_8$ and H$_0$ is largely broken, consistent with MCMC noise. Our analysis echoes the conclusion of earlier work, using previous datasets, that these tensions cannot be resolved in a flat universe with scalar perturbations and total neutrino mass of 0.06 $\eV$, where cold dark matter decays into dark radiation~\cite{Enqvist:2015ara,Poulin:2016nat,Chudaykin:2020aoj,Nygaard:2020sow}. We further conclude that these tensions cannot be resolved in extended DCDM models, as demonstrated in Table \ref{tab:ext_models}. 

In this work, we have focused on constraining the decay rate of dark matter from observations on linear scales. In the following, we touch briefly on the possibility of using observations at non-linear scales. We highlighted before the consistency of our results with Ref. \cite{Simon:2022ftd}, even when they include mildly non-linear scales in their analysis via the EFTofLSS framework. This could suggest that there is no significant improvement in the DCDM constraints by only going to mildly non-linear scales.
Further, ref. \cite{Hubert:2021khy} showed that DCDM leads to a notable suppression of power on highly non-linear scales. They took advantage of this information by using a fitting function derived from N-body simulations to assess the sensitivity of a Euclid-like weak lensing survey to the DCDM decay rate and fraction. Indeed, future surveys devoted to the observation of the large-scale structure of the Universe and its evolution at late times will detect signals on small scales with high sensitivity. In this regard, the analysis of Refs.~\cite{Hubert:2021khy,Simon:2022ftd} clearly points to the ability of these measurements to constrain scenarios in which the late-time behaviour of cosmological components deviates from the $\Lambda$CDM paradigm. Complemented with a more robust understanding of the physics of non-linear structure evolution, we expect future large-scale-structure surveys (e.g. Euclid~\cite{EUCLID:2011zbd}, DESI~\cite{DESI:2013agm}, and Rubin Observatory~\cite{LSSTScience:2009jmu}) to further improve over current constraints on the decaying dark matter model discussed in this work.
We also note that further improvements will come from the inclusion of CMB data at small angular scales, where the effect of gravitational lensing is more pronounced. In Figure~\ref{fig:CMB_Decay}, we reported the region $\ell>3000$ to highlight the possible improvement coming from observations at those scales, especially in polarization. This region of angular scales is not probed by Planck, but is probed by the ground-based experiments ACT~\cite{ACT:2020gnv,ACT:2020frw} and SPT~\cite{SPT-3G:2021wgf,SPT-3G:2021eoc}, and will be probed by the upcoming experiment Simons Observatory~\cite{SimonsObservatory:2018koc} and, later, by CMB-S4~\cite{Abazajian:2019eic}. The inclusion of this range of scales requires a more careful treatment of non-linear effects in the growth of cosmic structure, which we defer to future work. Future CMB experiments will also provide unprecedented measurements of the power spectrum of gravitational lensing over a wider range of angular scales than that currently covered. We expect further improvements in the constraints of DCDM parameters to come from this direction (and possible cross-correlations with direct probes of galaxy clustering) as well.

\acknowledgments
We acknowledge the financial support from the INFN InDark initiative and from the COSMOS network (www.cosmosnet.it) through the ASI (Italian Space Agency) grants 2016-24-H.0 and 2016-24-H.1-2018. TB was supported through the INFN project GRANT73/Tec-Nu.  We acknowledge the use of \texttt{GetDist} \citep{Lewis:2019xzd} software package, and the use of computational resources granted by Cineca.

\bibliographystyle{JHEP}
\bibliography{DCDM.bib,Planck_bib.bib}

\appendix
\section{Appendix}\label{sec:appA}
\begin{figure}[tbp]
\centering
\includegraphics[width=1.0\textwidth]{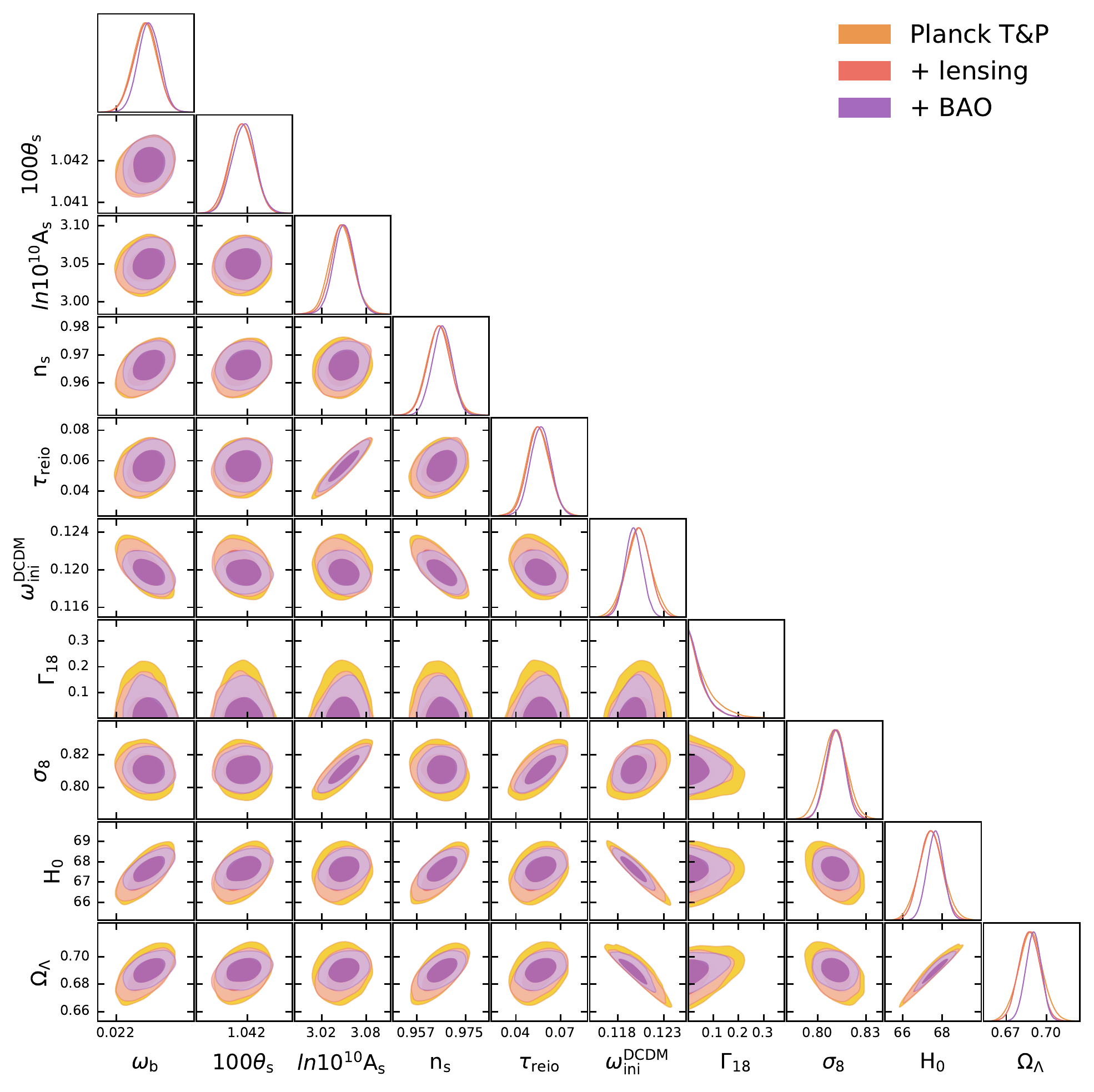}
\hfill
\caption{Constraints on parameters of the $\Lambda$CDM + $\Gamma_{18}$ model with $f_\mathrm{DCDM}=1$ (all dark matter is unstable) from the combination of Planck 2018 CMB temperature and polarization data alone (grey), with Planck 2018 CMB lensing power spectrum (red), and additionally with BAO measurements from eBOSS DR16 and earlier datasets (blue). For the dark matter decay rate, $\Gamma_{18}$, the lensing information makes the largest contribution to information from the primary anisotropies, while for the rest of the parameters the BAO data makes a bigger difference.}
\label{fig:gamma_degen}
\end{figure}

\begin{figure}[tbp]
	\centering
	\includegraphics[width=0.90\textwidth,clip]{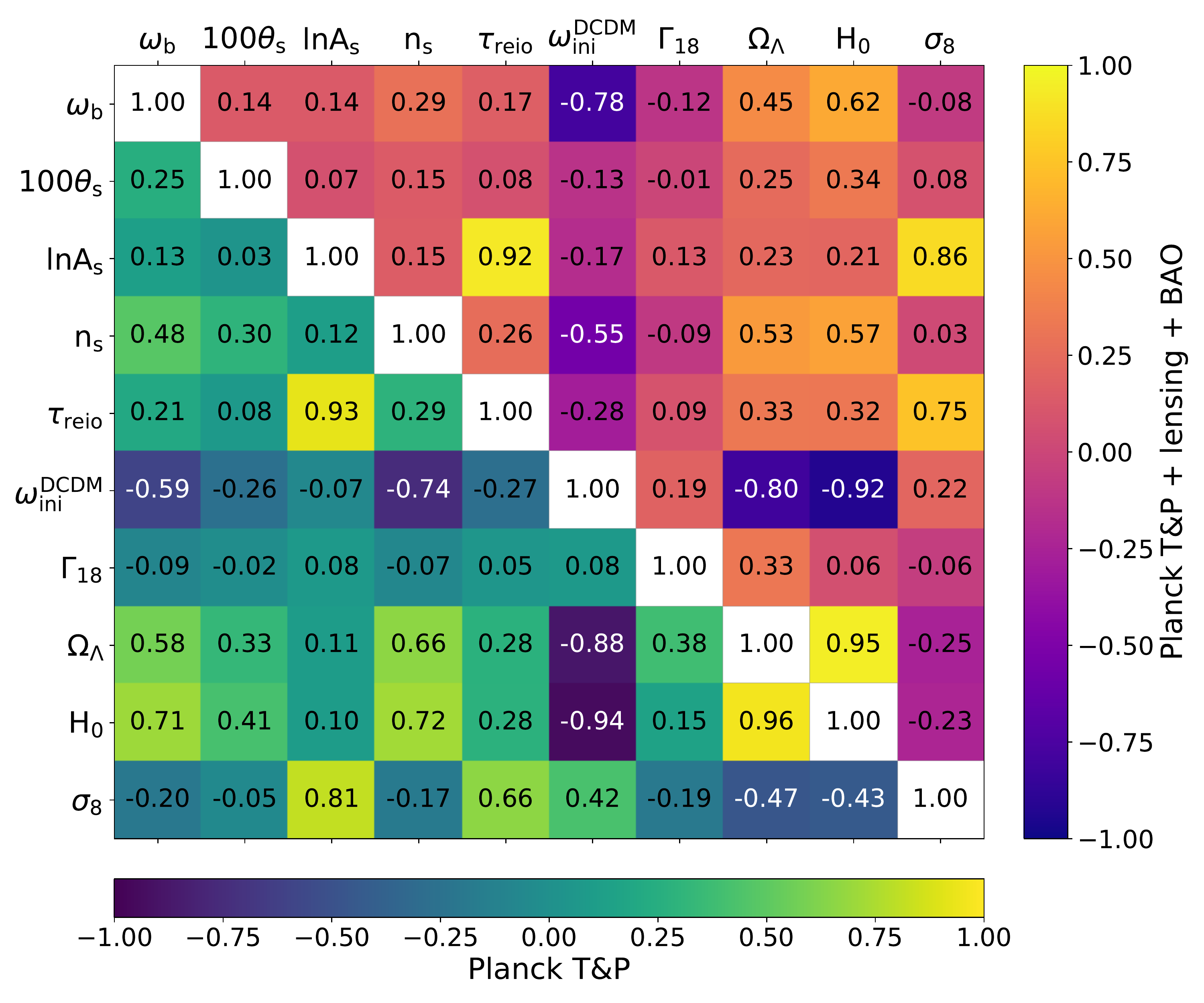}
	\hfill
	\caption{\label{fig:corr_matrix} Correlation matrix of the parameters of DCDM model with $f_\mathrm{DCDM}=1$ (all dark matter unstable) sampled in this analysis and a subset of derived parameters.
	The value of the correlation coefficient between any two parameters is written in the corresponding box. Larger coefficients in absolute value signal a greater level of correlation (and the sign shows the direction of correlation).
	In the lower diagonal (upper diagonal) part of the matrix we show the correlation coefficient for the Planck T\&P (Planck T\&P+lensing+BAO) dataset. The DM decay rate is mostly correlated with the initial dark matter abundance $\omega{}_\text{ini}^\text{DCDM}$, the $\Omega_\Lambda$ density parameter, the Hubble constant $H_0$ and the amplitude of matter perturbations at small scales $\sigma_8$.
	}
\end{figure}

We provide some additional plots and constraints not included in the main body of the paper. We focus on the case $f_\mathrm{DCDM}=1$, i.e., all the dark matter is unstable. In Table \ref{tab:all_param_derived}, we show the constraints on the full set of parameters in our model i.e. the standard $\Lambda$CDM and the decay rate of dark matter, $\Gamma_{18}$. Additionally, we also show derived constraints on the parameters H$_0$ and $\sigma_8$ which have been the source of tension in modern cosmology. We report the 68\% confidence interval for parameter which have both the upper and lower limit and the 95\% C.L. limits for parameters that are only upper or lower bounded. We show degeneracies between parameters for the different datasets considered in this work (see Section~\ref{sec:data_param_est} for more details) in Figure~\ref{fig:gamma_degen}. To visualize the degeneracy between the parameters in another way, we show in Figure \ref{fig:corr_matrix} the correlation matrix for the same parameters. The lower triangle of Figure \ref{fig:corr_matrix} is derived from Planck T\&P and the upper triangle from Planck T\&P plus CMB lensing and BAO data.  

\begin{table}[h]
\centering
{\fontsize{11pt}{18pt}\selectfont
\begin{tabular}{l|l|l|l}
\hline
\textbf{Parameter\textbackslash Dataset}         & \textbf{Planck T\&P}                                                           & \textbf{+ lensing}                                                            & \textbf{+ BAO}                                                                \\ \hline\hline
$\omega_\text{b}$ & 0.02235 $_{- 0.00016}^{+ 0.00015}$  & 0.02236  $_{- 0.00014}^{+ 0.00016}$  & 0.02241  $_{- 0.00012}^{+ 0.00014}$                                                \\ \hline
$100\theta{}_\text{s}$ & 1.04187 $\pm0.00030$ & 1.04186 $_{- 0.00030}^{+ 0.00033}$    & 1.04191 $_{- 0.00030}^{+ 0.00029}$ \\ \hline
$\ln 10^{10}A_\text{s }$           & 3.046 $_{- 0.017}^{+ 0.016}$                                                     & 3.048 $\pm$ 0.015                                                              & 3.051 $_{- 0.014}^{+ 0.013}$                                                    \\ \hline
n$_\text{s}$                       & 0.9652 $_{- 0.0045}^{+ 0.0046}$                                                            & 0.9652 $_{- 0.0043}^{+ 0.0042}$                                                  & 0.9665 $_{- 0.0035}^{+ 0.0039}$                                                           \\ \hline
$\tau_\text{reio}$                 & 0.0546 $_{- 0.0082}^{+ 0.0077}$                                                  & 0.0552 $_{- 0.0077}^{+ 0.0076}$                                                 & 0.0570 $_{- 0.0075}^{+ 0.0064}$                                                 \\ \hline
$\omega{}_\text{ini}^\text{DCDM }$ & 0.1203 $_{- 0.0013} ^{+ 0.0015}$                                                            & 0.1202 $\pm$ 0.0013                                                & 0.11967 $_{- 0.00092}^{+ 0.00096}$                                              \\ \hline
$\Gamma_{18} = \Gammadm/10^{-18}\,\mathrm{s}^{-1}$ & \begin{tabular}[c]{@{}l@{}}$<0.175$ (95\%)\end{tabular}     & \begin{tabular}[c]{@{}l@{}} $<0.136$ (95\%)\end{tabular}   & \begin{tabular}[c]{@{}l@{}} $<0.129$ (95\%)\end{tabular}             \\ \hline
$\tau_\text{DCDCM}$                & \begin{tabular}[c]{@{}l@{}} $>181$ (95\%)\end{tabular} & \begin{tabular}[c]{@{}l@{}} $>234$ (95\%)\end{tabular} & \begin{tabular}[c]{@{}l@{}} $>246$ (95\%)\end{tabular} \\ \hline
$\sigma_8$                         & 0.8102 $_{- 0.0080}^{+ 0.0073}$                                                  & 0.8113 $_{- 0.0061}^{+ 0.0064}$                                                           & 0.8110 $_{- 0.0062}^{+ 0.0056}$                                                           \\ \hline
H$_0$                       & 67.44 $\pm$ 0.63                                                       & 67.42 $_{- 0.58}^{+ 0.56}$                                                      & 67.67 $_{- 0.44}^{+ 0.39}$                                                      \\ \hline
\end{tabular}}
\caption{Constraints for the five of the standard cosmological parameters, the DCDM model parameters ($\omega{}_\text{ini}^\text{DCDM }$, $\Gamma_{18 }$) and some interesting derived parameters, for $f_\mathrm{DCDM}=1$ (all dark matter unstable), for the three combinations of datasets considered in this work (see Section~\ref{sec:data_param_est} for details). All bounds are 68\% confidence intervals unless otherwise indicated.}
\label{tab:all_param_derived}
\end{table}

\begin{figure}[tbp]
	\centering
	\includegraphics[width=0.90\textwidth,clip]{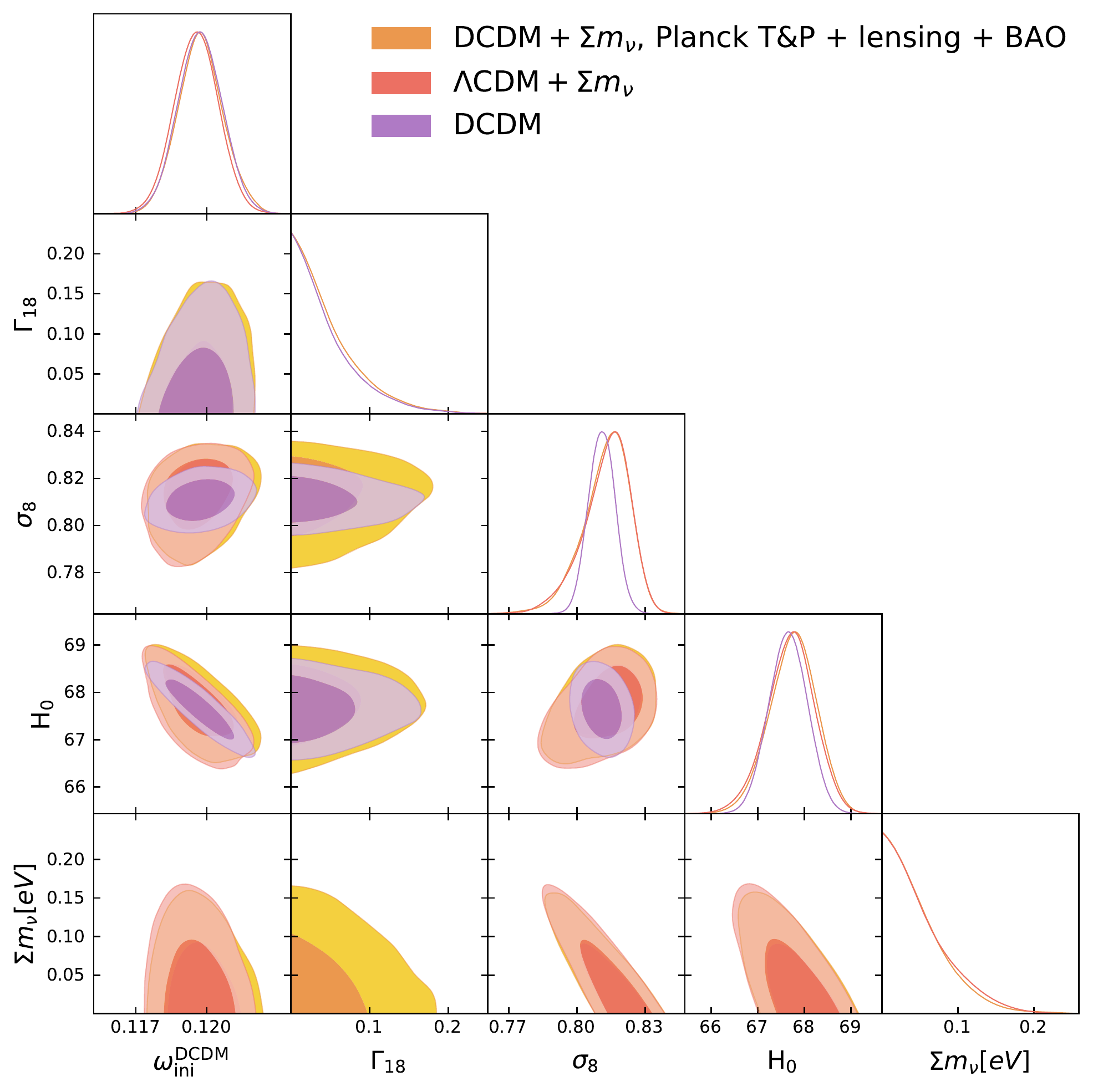}
	\hfill
	\caption{\label{fig:muLDDM+LCDM}Posterior distributions for the our baseline case along with the posteriors for the sum of neutrino mass. One can appreciate in the plot that the DCDM decay rate and sum of neutrino masses are almost uncorrelated. This is also reflected in the 1D marginalized of $\Sigma m_\nu$ and $\Gamma_{18}$ where it can been that posterior for $\Gamma_{18}$ in the extended model nearly overlaps with our baseline case. Similarly, 1D posterior for $\Sigma m_\nu$ overlaps with the those obtained in $\Lambda$CDM universe. We point out that in the $\Lambda$CDM model, $\omega^\text{DCDM}_\text{ini} = \omega_\text{CDM}$.}
\end{figure}

\begin{figure}[tbp]
	\centering
	\includegraphics[width=0.90\textwidth,clip]{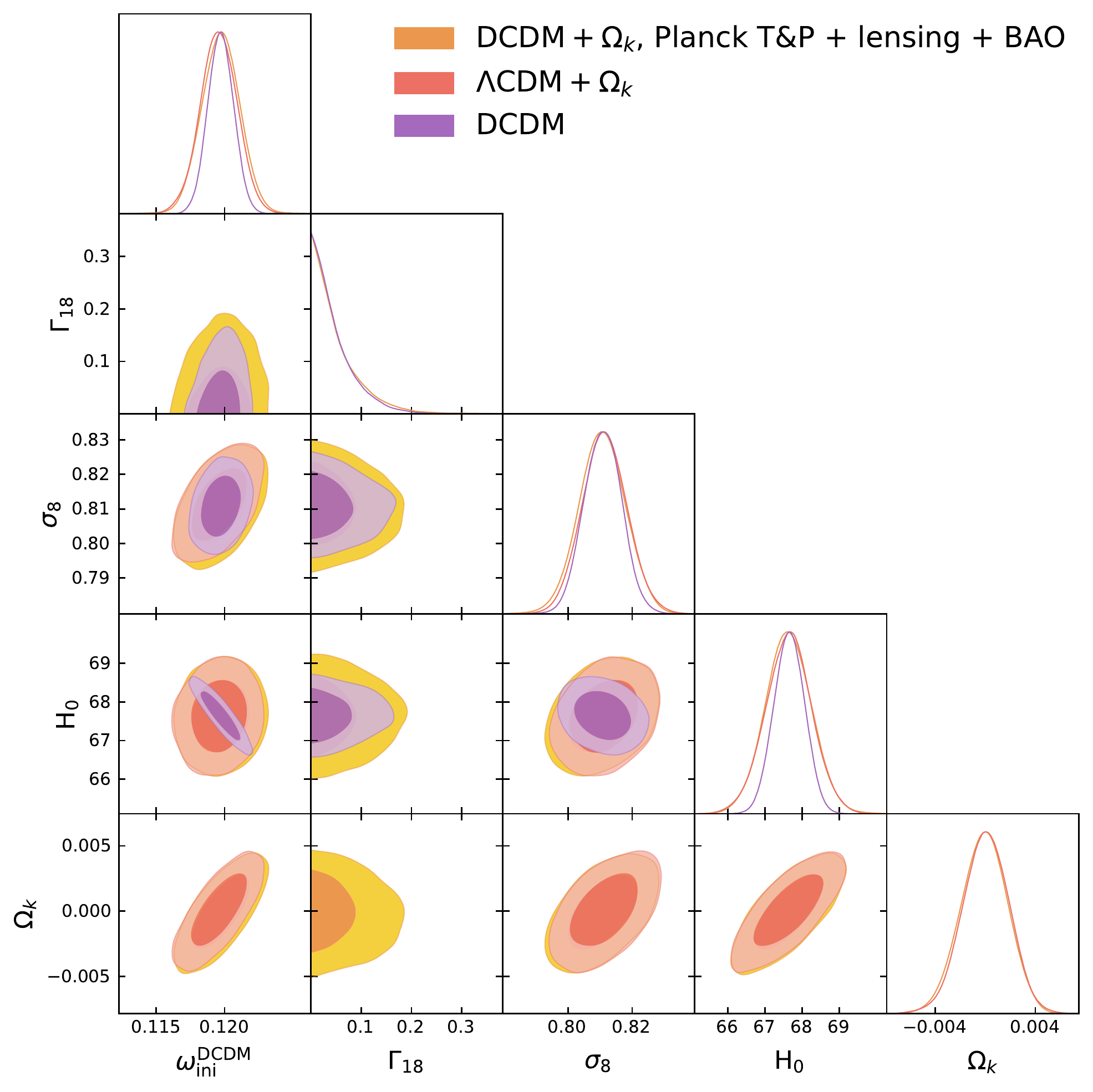}
	\hfill
	\caption{\label{fig:OkLDDM+LCDM}Posterior distributions for the our baseline case along with the posteriors for the curvature density parameter. One can appreciate in the plot that the DCDM decay rate and curvature density parameter are almost uncorrelated. This is also reflected in the 1D marginalized of $\Omega_K$ and $\Gamma_{18}$ where it can been that posterior for $\Omega_{K}$ in the extended model nearly overlaps with our baseline case. Similarly, 1D posterior for $\Omega_K$ overlaps with the those obtained in $\Lambda$CDM universe. We point out that in the $\Lambda$CDM model, $\omega^\text{DCDM}_\text{ini} = \omega_\text{CDM}$.}
\end{figure}

\begin{figure}[tbp]
	\centering
	\includegraphics[width=0.90\textwidth,clip]{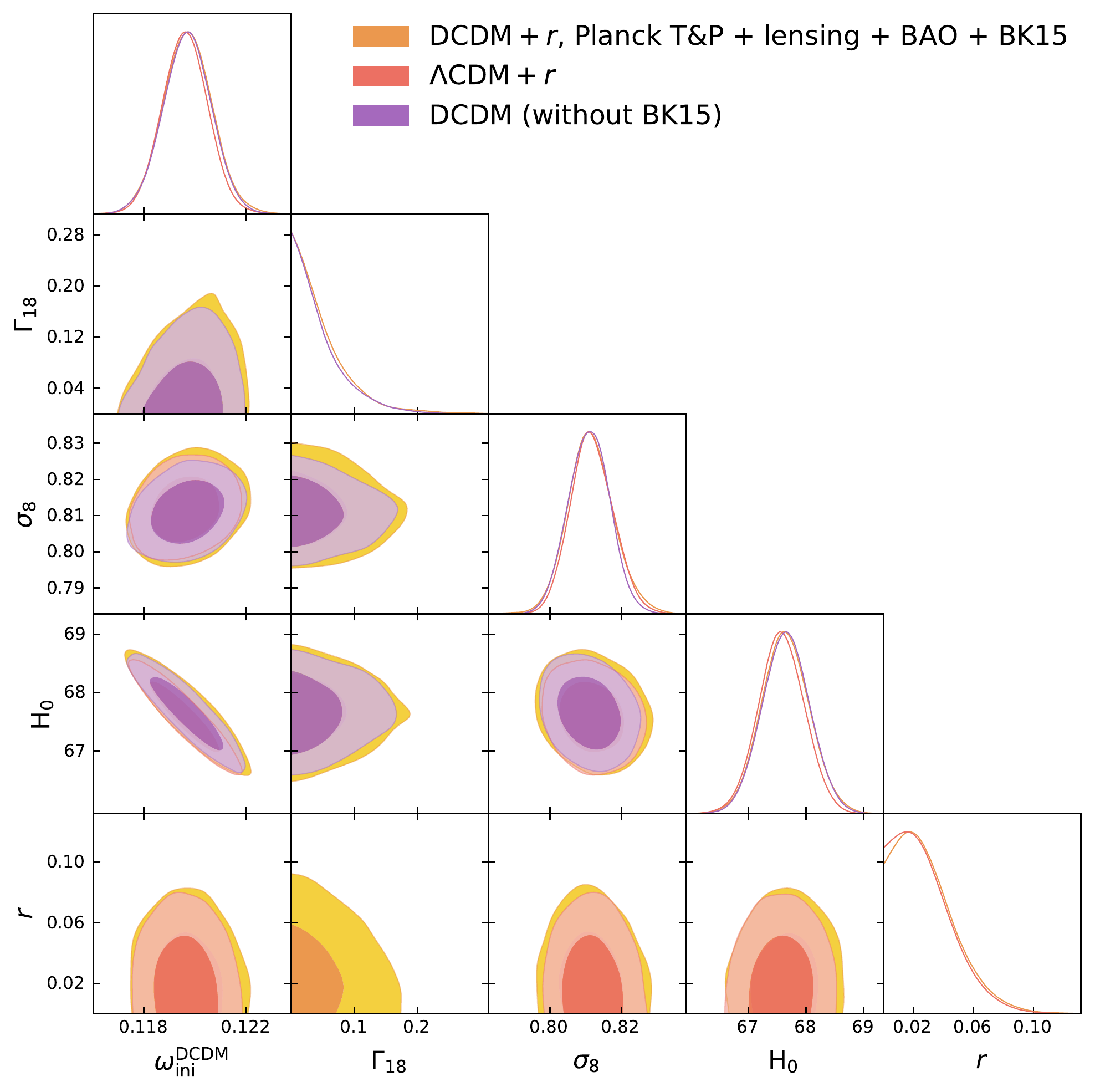}
	\hfill
	\caption{\label{fig:rLDDM+LCDM}Posterior distributions for the our baseline case along with the posteriors for the tensor-to-scalar ratio, r. One can appreciate in the plot that the DCDM decay rate and tensor-t-scalar ratio are almost uncorrelated. One can appreciate that the 1D posterior of $\Gamma_{18}$ are almost identical to those obtained in our baseline DCDM model. The 1D posterior of r show that in the DCDM model, smaller values of r are more tightly constrained. We point out that in the $\Lambda$CDM model, $\omega^\text{DCDM}_\text{ini} = \omega_\text{CDM}$.}
\end{figure}
\end{document}